\documentclass[pdflatex,sn-basic]{sn-jnl}

\usepackage{xcolor}
\usepackage{physics}
\usepackage{manyfoot}
\usepackage{amssymb}
\usepackage{siunitx}[=v2]
\DeclareSIUnit{\erg}{erg}
\DeclareSIUnit{\Mass}{\mathit{M}}
\DeclareSIUnit{\c}{\mathit{c}}
\DeclareSIUnit{\dyn}{dyn}
\DeclareSIUnit{\Gauss}{G}
\DeclareSIQualifier{\Sun}{\ensuremath{\odot}}
\DeclareSIQualifier{\disk}{disk}

\usepackage{comment}

\usepackage{glossaries}
\newcommand{\ave}[1]{\left\langle{#1}\right\rangle}
\newcommand{\fluct}[1]{\delta {#1}}

\usepackage{bm}
\renewcommand{\vec}[1]{\bm{\mathrm{#1}}}
\newcommand{\BB}{\vec{B}}
\newcommand{\nn}{\vec{n}}
\newcommand{\vv}{\vec{v}}
\newcommand{\VV}{\vec{v}}

\newcommand{\sect}{Sect.}
\newcommand{\sects}{Sects.}
\newcommand{\fig}{Fig.}
\newcommand{\figs}{Figs.}
\newcommand{\eqn}{Eq.}
\newcommand{\eqns}{Eqs.}

\hypersetup{allcolors=black}

\usepackage{acronym}
\acrodef{ADM}{Arnowitt-Deser-Misner}
\acrodef{AMR}{adaptive mesh-refinement}
\acrodef{BH}{black hole}
\acrodef{BBH}{binary black-hole}
\acrodef{BHNS}{black-hole neutron-star}
\acrodef{BNS}{binary neutron star}
\acrodef{CCSN}{core-collapse supernova}
\acrodefplural{CCSN}[CCSNe]{core-collapse supernovae}
\acrodef{DG}{discontinuous Galerkin}
\acrodef{DNS}{direct numerical simulation}
\acrodef{EM}{electromagnetic}
\acrodef{ET}{Einstein Telescope}
\acrodef{EOB}{effective-one-body}
\acrodef{EOS}{equation of state}
\acrodef{GR}{general relativistic}
\acrodef{GRHD}{general-relativistic hydrodynamics}
\acrodef{GRLES}{general-relativistic large-eddy simulation}
\acrodef{GRMHD}{general-relativistic magnetohydrodynamics}
\acrodef{GW}{gravitational wave}
\acrodef{HD}{hydrodynamics}
\acrodef{HRSC}{high-resolution shock capturing}
\acrodef{KH}{Kelvin Helmholtz}
\acrodef{KHI}{Kelvin Helmholtz instability}
\acrodef{ILES}{implicit large-eddy simulation}
\acrodef{ISCO}{innermost circular orbit}
\acrodef{LES}{large-eddy simulation}
\acrodef{MAD}{magnetically arrested disk}
\acrodef{MHD}{magnetohydrodynamics}
\acrodef{RMNS}{remnant massive neutron star}
\acrodef{MRI}{magneotorational instability}
\acrodef{NR}{numerical relativity}
\acrodef{NS}{neutron star}
\acrodef{RANS}{Reynolds-Averaged Navier-Stokes}
\acrodef{SGRB}{short gamma-ray burst}
\acrodef{SPH}{smoothed-particle hydrodynamics}
\acrodef{SN}{supernova}
\acrodef{SNR}{signal-to-noise ratio}

\makeglossaries

\newglossaryentry{Dt}{
description={The material derivative $D_t = \partial_t + \vv \cdot \nabla$, where $\vv$ is the velocity},
name={\ensuremath{D_t}},
sort=D
}

\newglossaryentry{B}{
name={\ensuremath{\BB}},
description={Magnetic field},
sort=B
}
\newglossaryentry{Bmag}{
name={\ensuremath{B}},
description={Magnetic field strength or magnitude},
sort=B
}

\newglossaryentry{3vel}{
name={\ensuremath{\vv}},
description={Newtonian 3-velocity},
sort=vel
}
\newglossaryentry{3velmag}{
name={\ensuremath{v}},
description={Magnitude of Newtonian 3-velocity},
sort=vel
}
\newglossaryentry{nu}{
name={\ensuremath{\nu}},
description={Kinematic viscosity, assumed constant},
sort=nu
}
\newglossaryentry{rho}{
name={\ensuremath{\rho}},
description={Matter density},
sort=rho
}
\newglossaryentry{pressure}{
name={\ensuremath{p}},
description={Isotropic pressure},
sort=pressure
}
\newglossaryentry{Omega}{
name={\ensuremath{V}},
description={Local volume},
sort=Volume
}
\newglossaryentry{normal}{
name={\ensuremath{\nn}},
description={Normal to local volume \gls{Omega}},
sort=normal
}
\newglossaryentry{lengthscale}{
name={\ensuremath{\ell}},
description={Characteristic lengthscale},
sort=lengthscale
}
\newglossaryentry{ld}{
name={\ensuremath{\ell_d}},
description={Dissipation lengthscale},
sort=lengthscaled
}
\newglossaryentry{le}{
name={\ensuremath{\ell_e}},
description={Effective electron dissipation lengthscale},
sort=lengthscalee
}
\newglossaryentry{lB}{
name={\ensuremath{\ell_B}},
description={Magnetic dissipation lengthscale},
sort=lengthscaleB
}
\newglossaryentry{l0}{
name={\ensuremath{\ell_0}},
description={Outer lengthscale of inertial range},
sort=lengthscalezero
}
\newglossaryentry{lmix}{
name={\ensuremath{\ell_{\rm mix}}},
description={Mixing lengthscale},
sort=lengthscalemix
}
\newglossaryentry{H}{
name={\ensuremath{H}},
description={Disk thickness},
sort=H
}
\newglossaryentry{Re}{
name={\ensuremath{\mathrm{Re}}},
description={Reynolds number, a dimensionless number quantifying the impact of viscosity},
sort=Reynolds
}
\newglossaryentry{nu_neutrino}{
name={\ensuremath{\nu_\nu}},
description={Kinematic viscosity due to neutrinos},
sort=nun
}
\newglossaryentry{nu_e}{
name={\ensuremath{\nu_e}},
description={Kinematic viscosity due to electrons},
sort=nue
}
\newglossaryentry{nu_T}{
name={\ensuremath{\nu_T}},
description={Turbulent (or eddy) viscosity in the Boussinesq hypothesis},
sort=nuT
}
\newglossaryentry{nu_eff}{
name={\ensuremath{\nu_{\mathrm{eff}}}},
description={Effective viscosity in closure models},
sort=nueff
}
\newglossaryentry{T}{
name={\ensuremath{T}},
description={Temperature},
sort=Temperature
}
\newglossaryentry{Yp}{
name={\ensuremath{Y_p}},
description={Proton fraction},
sort=Yp
}
\newglossaryentry{np}{
name={\ensuremath{n_p}},
description={Proton number},
sort=np
}
\newglossaryentry{nn}{
name={\ensuremath{n_n}},
description={Neutron number},
sort=nn
}
\newglossaryentry{lnu}{
name={\ensuremath{\ell_\nu}},
description={Dissipation lengthscale due to neutrinos},
sort=lengthscalenu
}
\newglossaryentry{lambdanu}{
name={\ensuremath{\lambda_\nu}},
description={Mean-free path of neutrinos},
sort=lambda
}
\newglossaryentry{eta}{
name={\ensuremath{\eta}},
description={Magnetic resistivity},
sort=eta
}
\newglossaryentry{beta}{
name={\ensuremath{\beta}},
description={Plasma $\beta$ parameter, $\beta = \gls{pressure} / (\gls{Bmag}^2 / 8 \pi)$},
sort=beta
}
\newglossaryentry{e}{
name={\ensuremath{e}},
description={Relativistic energy density},
sort=energy
}
\newglossaryentry{u}{
name={\ensuremath{u^a}},
description={Relativistic 4-velocity},
sort=u
}
\newglossaryentry{n}{
name={\ensuremath{n}},
description={Particle number},
sort=n
}
\newglossaryentry{ndot}{
name={\ensuremath{\dot{n}}},
description={Material derivative of particle number, $\dot{n} = \gls{u} \nabla_a \gls{n}$},
sort=ndot
}
\newglossaryentry{n_current}{
name={\ensuremath{n^a}},
description={Particle number current $n^a = \gls{n} \gls{u}$},
sort=n
}
\newglossaryentry{stress-energy}{
name={\ensuremath{T^a_b}},
description={Relativistic stress-energy tensor},
sort=T
}
\newglossaryentry{additional}{
name={\ensuremath{A^a_b}},
description={Additional part of the mean stress-energy},
sort=A
}
\newglossaryentry{dissip}{
name={\ensuremath{D^a_b}},
description={Dissipative part of the full stress-energy as seen by the averaged flow},
sort=Dissip
}
\newglossaryentry{residual-e}{
name={\ensuremath{R_e}},
description={Residual part of the relativistic total energy as seen by the averaged flow},
sort=Re
}
\newglossaryentry{residual-p}{
name={\ensuremath{R_p}},
description={Residual part of the relativistic total pressure as seen by the averaged flow},
sort=Rp
}
\newglossaryentry{heat}{
name={\ensuremath{q^a}},
description={(Effective) relativistic heat flux},
sort=q
}
\newglossaryentry{bulk-viscous}{
name={\ensuremath{\Pi}},
description={(Effective) relativistic bulk viscous pressure},
sort=Pi
}
\newglossaryentry{shear-viscous}{
name={\ensuremath{\pi^a_b}},
description={(Effective) relativistic shear stress},
sort=pi
}
\newglossaryentry{chemical-potential}{
name={\ensuremath{\mu}},
description={Chemical potential},
sort=mu
}
\newglossaryentry{passive-scalar}{
name={\ensuremath{\phi}},
description={Generic passive scalar},
sort=phi
}
\newglossaryentry{anisotropic-stress}{
name={\ensuremath{a^i_j}},
description={Anisotropic part of the Newtonian Reynolds stress},
sort=anisotropic
}
\newglossaryentry{tke}{
name={\ensuremath{k_{\mathrm{TKE}}}},
description={Turbulent kinetic energy},
sort=k
}
\newglossaryentry{newtonian-energy}{
name={\ensuremath{E}},
description={Newtonian specific energy},
sort=Energy
}
\newglossaryentry{ke-flux}{
name={\ensuremath{\bm{T}}},
description={Newtonian kinetic energy flux},
sort=T
}
\newglossaryentry{rate-of-strain}{
name={\ensuremath{S}},
description={Newtonian rate of strain tensor},
sort=Strain
}
\newglossaryentry{production}{
name={\ensuremath{\mathcal{P}}},
description={Newtonian production of turbulent kinetic energy},
sort=Production
}
\newglossaryentry{tke-dissipation}{
name={\ensuremath{\varepsilon}},
description={Newtonian dissipation of turbulent kinetic energy},
sort=epsilon
}
\newglossaryentry{tetrad}{
name={\ensuremath{e^b_{(j)}}},
description={An orthonormal tetrad},
sort=e
}
\newglossaryentry{k-wave}{
name={\ensuremath{k}},
description={Wavenumber},
sort=k
}
\newglossaryentry{j-specific}{
name={\ensuremath{j}},
description={Specific angular momentum},
sort=jspecific
}
\newglossaryentry{J-total}{
name={\ensuremath{J}},
description={Total angular momentum},
sort=J
}
\newglossaryentry{J-total-dot}{
name={\ensuremath{\dot{J}}},
description={Rate of change of total angular momentum \gls{J-total}},
sort=Jdot
}
\newglossaryentry{Mb-total}{
name={\ensuremath{M_b}},
description={Total baryon mass},
sort=M
}
\newglossaryentry{alpha-viscosity}{
name={\ensuremath{\alpha}},
description={alpha viscosity},
sort=alpha
}
\newglossaryentry{dx}{
name={\ensuremath{\Delta x}},
description={Numerical grid spacing},
sort=d
}
\newglossaryentry{v0}{
name={\ensuremath{v_0}},
description={Characteristic velocity scale},
sort=vel
}
\newglossaryentry{tau_rs}{
name={\ensuremath{\tau^a_b}},
description={Residual (non-ideal) stress-energy after averaging or filtering},
sort=tau
}
\newglossaryentry{s}{
name={\ensuremath{s}},
description={Entropy},
sort=entropy
}
\newglossaryentry{creation_rate}{
name={\ensuremath{\Gamma_n}},
description={Particle creation rate},
sort=Gamma
}
\newglossaryentry{perp}{
name={\ensuremath{\perp^c_a}},
description={Projector into spatial slice $\perp^c_a = \delta^c_a + \glslink{N_slice}{N^c} \glslink{N_slice}{N_a}$, where $\glslink{N_slice}{N^a}$ is the normal to the slice},
sort=Perp
}
\newglossaryentry{vbar}{
name={\ensuremath{\bar{v}_c}},
description={3-velocity of the mean flow in the spatial slice, $\glslink{perp}{\perp^d_c} \ave{\glslink{u}{u_d}}$},
sort=v
}
\newglossaryentry{gamma_metric}{
name={\ensuremath{\gamma_{cd}}},
description={Spatial 3-metric $\gamma_{cd} = \glslink{perp}{\perp^a_c} \glslink{perp}{\perp^b_d} g_{ab}$},
sort=gamma
}
\newglossaryentry{N_slice}{
name={\ensuremath{N^c}},
description={Normal to spatial slice},
sort=N
}
\newglossaryentry{cs}{
name={\ensuremath{c_s}},
description={Speed of sound},
sort={speed}
}
\newglossaryentry{Omega_ang}{
name={\ensuremath{\Omega}},
description={Angular velocity},
sort=Angular
}
\newglossaryentry{J_current}{
name={\ensuremath{J^a}},
description={Mass current $J^a = m \gls{n_current}$},
sort=Jmass
}
\newglossaryentry{D}{
name={\ensuremath{D}},
description={Conserved mass seen by the normal observer, $\gls{J_current} \glslink{N_slice}{N_a}$},
sort=D
}
\newglossaryentry{E}{
name={\ensuremath{E}},
description={Energy density as viewed by the normal observer, $\gls{stress-energy}\glslink{N_slice}{N_a}\glslink{N_slice}{N^b}$},
sort=E
}
\newglossaryentry{Si}{
name={\ensuremath{S_i}},
description={Linear momentum density as viewed by the normal observer, $\glslink{N_slice}{N_a}\glslink{perp}{\perp^b_i}\gls{stress-energy}$},
sort=Si
}
\newglossaryentry{Sij}{
name={\ensuremath{S_{ij}}},
description={Stress tensor as viewed by the normal observer, $\glslink{perp}{\perp_{ai}}\glslink{perp}{\perp^b_j}\gls{stress-energy}$},
sort=Sij
}
\newglossaryentry{alpha_3metric}{
name={\ensuremath{\alpha}},
description={Lapse function from the normal to the slice, $-\glslink{N_slice}{N_0}$},
sort=alpha
}
\newglossaryentry{gamma}{
name={\ensuremath{\gamma}},
description={Determinant of the 3-metric \gls{gamma_metric}},
sort=gamma
}
\newglossaryentry{g}{
name={\ensuremath{g}},
description={Determinant of the 4-metric $g_{ab}$},
sort=g
}
\newglossaryentry{Wlorentz}{
name={\ensuremath{W}},
description={Lorentz factor $W = -\gls{u} \glslink{N_slice}{N_a}$},
sort=W
}
\newglossaryentry{r3vel}{
name={\ensuremath{v^a}},
description={Relativistic 3-velocity $v^a = \glslink{perp}{\perp^a_b} \glslink{u}{u^b} / \gls{Wlorentz}$},
sort=velocity
}
\newglossaryentry{observer}{
name={\ensuremath{U^a}},
description={4-velocity of an observer},
sort=U
}

\begin{document}

\title[Turbulence]{Turbulence modelling in neutron star merger simulations}

\author*[1,2,3]{\fnm{David} \sur{Radice}}\email{dur566@psu.edu}

\author[4]{\fnm{Ian} \sur{Hawke}}\email{I.Hawke@soton.ac.uk}
\equalcont{These authors contributed equally to this work.}

\affil*[1]{\orgdiv{Institute for Gravitation \& the Cosmos}, \orgname{The Pennsylvania State University}, \orgaddress{\street{ University Park}, \city{PA}, \postcode{16802}, \country{USA}}}

\affil[2]{\orgdiv{Department of Physics}, \orgname{The Pennsylvania State University}, \orgaddress{\street{ University Park}, \city{PA}, \postcode{16802}, \country{USA}}}

\affil[3]{\orgdiv{Department of Astronomy \& Astrophysics}, \orgname{The Pennsylvania State University}, \orgaddress{\street{ University Park}, \city{PA}, \postcode{16802}, \country{USA}}}

\affil[4]{\orgdiv{Mathematical Sciences and STAG Research Centre}, \orgname{University of Southampton}, \orgaddress{\city{Southampton}, \postcode{SO17 1BJ}, \country{United Kingdom}}}


\abstract{Observations of neutron star mergers have the potential to unveil detailed physics of matter and gravity in regimes inaccessible by other experiments. Quantitative comparisons to theory and parameter estimation require nonlinear numerical simulations. However, the detailed physics of energy and momentum transfer between different scales, and the formation and interaction of small scale structures, which can be probed by detectors, are not captured by current simulations. This is where turbulence enters neutron star modelling. This review will outline the theory and current status of turbulence modelling for relativistic neutron star merger simulations.}



\maketitle

\clearpage
\renewcommand{\glossarypreamble}{\glsfindwidesttoplevelname[\currentglossary]}
\setglossarystyle{alttree}
\printglossaries
\clearpage

\renewcommand{\sectionmark}[1]{\markboth{#1}{#1}}

\section{Introduction}\label{sec:intro}

\Ac{NS} mergers, including both \ac{BNS} and \ac{BHNS} mergers, are connected to some of the most pressing open questions in nuclear and high-energy astrophysics, such as the origin of elements heavier than iron, the nature of matter at supernuclear densities, and the mechanism producing short-gamma ray bursts \citep{baiotti_binary_2017}. Multi-messenger observations of \ac{NS} mergers enabled by ground-based \ac{GW} detectors LIGO, Virgo, and KAGRA have revolutionised our understanding of these events \citep{2018PhRvL.121p1101A,2019PhRvX...9a1001A}. However, quantitative theoretical frameworks are needed to turn these extraordinary observations into insight. Here, \ac{NR} has a key role. \ac{NR} simulations have enabled the development of accurate \ac{GW} waveform models that have been used to constrain the tidal response of \acp{NS} \citep{de_tidal_2018,2019PhRvX...9a1001A,2018PhRvL.121p1101A}. 
\ac{NR} simulations have also been instrumental for the understanding of the \ac{EM} emission from merging \acp{NS} \citep{hotokezaka_mass_2013, bauswein_systematics_2013, paschalidis_relativistic_2015, kyutoku_dynamical_2015, sekiguchi_dynamical_2015, foucart_dynamical_2017, lehner_unequal_2016, ruiz_binary_2016, sekiguchi_dynamical_2016, radice_binary_2018, fujibayashi_mass_2018, nedora_spiral-wave_2019, vincent_unequal_2020, miller_full_2019, ciolfi_binary_2020, foucart_monte-carlo_2020, nedora_numerical_2021, combi_jets_2023, kiuchi_self-consistent_2023, kiuchi_large-scale_2023}. They have enabled joint \ac{EM} and \ac{GW} inference providing tight constraints on the \ac{EOS} of \acp{NS} and on the r-process nucleosynthesis yields of the mergers \citep{margalit_constraining_2017, bauswein_neutron-star_2017, radice_gw170817_2018, dietrich_multimessenger_2020, breschi_constraints_2022, ruiz_multimessenger_2021, rosswog_heavy_2022}.

Despite the many successes and the rapid progress in the last twenty years, \ac{NS} merger simulations still face huge challenges. Models of tidally interacting \acp{NS} with an order of magnitude smaller systematic errors will be required in the coming years \citep{gamba_waveform_2021}, as current \ac{GW} detectors improve in sensitivity and with the advent of next-generation ground-based \ac{GW} detectors, such as LIGO Voyager \citep{berti_snowmass2021_2022}, NEMO \citep{ackley_neutron_2020}, the Einstein Telescope \citep{punturo_einstein_2010}, and Cosmic Explorer \citep{reitze_cosmic_2019}. Quantitative models of the postmerger signal from \ac{BNS} mergers will also be required. Calculations of the \ac{EM} counterparts and of the nucleosynthesis yields from these systems need to be dramatically improved to match the expected reduction in the uncertainties in the rates and in the physics of neutron rich nuclei, which are expected from future observational campaigns and upcoming laboratory experiments \citep{zappa_binary_2023, schatz_horizons_2022}. A major obstacle is that simulations not only need to include more sophisticated microphysics, but they also need to be able to capture the chaotic, turbulent, \ac{MHD}\acused{GRMHD} flows that develop after the merger \citep{radice_dynamics_2020}.

In the context of \ac{BNS} mergers, hydrodynamic effects become dominant during the last orbit of the two stars \citep{radice_dynamics_2020}. In the case of binaries with mass ratio $q = M_2/M_1 \lesssim 0.75$ the least massive star is totally or partially tidally disrupted and accretes on the primary \citep{bernuzzi_accretion-induced_2020, perego_probing_2022}. At higher mass ratios (more symmetric binaries), the stars remain gravitationally bound and impact each other with radial velocities of the order of $\SI{0.1}{\c}$. At the time of contact or tidal disruption, the orbital velocity of the two centers of mass approaches ${\sim} \SI{0.2}{\c}$ \citep{radice_dynamics_2020}. The resulting relativistic shear flow is \ac{KH}\acused{KHI} unstable and becomes turbulent \citep{price_producing_2006}. The cores of the two stars bounce against each other, maybe multiple times, before finally merging to form a massive NS remnant (RMNS)\acused{RMNS}, or collapsing to form a \ac{BH}. In the process, they produce shock waves that accelerate a few $\SI{e-3}{\Mass\Sun}$ of neutron rich matter to escape velocity. They also squeeze material out of the remnant that was initially in their contact interface and that was previously stirred by the \ac{KH}. The expelled matter settles in an accretion disk with mass of up to a few $\SI{e-2}{\Mass\Sun}$ \citep{radice_binary_2018}. In the context of \ac{BHNS} mergers, if the \ac{NS} is disrupted outside of the \ac{ISCO} of the \ac{BH}, the part of the material that is not accreted and not gravitationally unbound eventually circularises to form a massive accretion disk \citep{kyutoku_coalescence_2021}. Shocks and strong shear flows are generated due to the tidal disruption and the self-intersection of the tidal streams.

In both \ac{BNS} and \ac{BHNS} mergers, the immediate outcome of the coalescence of the binary components is the formation of a compact object, a \ac{BH} or a massive \ac{NS}, surrounded by a turbulent accretion disk. Turbulence in the disk is also driven by the \ac{MRI} \citep{balbus_instability_1998}. The redistribution of angular momentum operated by turbulence determines the subsequent evolution of these systems, their multi-messenger emissions, and their nucleosynthetic yields. As we argue in \sect~\ref{sec:ns_conditions}, direct numerical simulations of such systems are impossible, due to the enormous separation of scales in these flows. Moreover, there are unsolved open questions concerning the nature of turbulence in hot, dense matter that need to be addressed. 

So far we have loosely used \emph{turbulence} in two senses: as a description of how a fluid transfers energy and momentum between different scales, and as a description of how the fluid processes form ordered structures and flows from disordered behaviour. The key features that can be demonstrated for simple models (such as the incompressible non-relativistic Navier-Stokes equations) include the coupling of power between wavenumbers until the physical viscosity exponentially damps the power, and the universal nature of the Kolmogorov cascade showing the power decaying with the wavenumber as $\gls{k-wave}^{-5/3}$. This has been backed up by experimental and numerical studies in Newtonian and relativistic~\citep{radice_universality_2013} cases. These points will be expanded in \sects~\ref{sec:ns_conditions}-\ref{sec:modelling}.

The implications for numerical modelling are profound. If the numerics can resolve the large wavenumber regime where viscosity acts then all the physical consequences of the theory can be captured. This is the regime of \ac{DNS}. However, if the numerics can only capture part of the wavenumber range where the power is transferred between scales, and not capture the high wavenumber range, then the numerical approximations will interfere with the physical consequences at high wavenumbers and, through the couplings inherent in the nonlinear terms, eventually on all scales. Correcting for these numerical limitations is the regime of \ac{LES}.

There are many physical and numerical reasons why more realistic models of neutron stars have more complex mechanisms for transferring energy and momentum between scales. In all cases, however, the fundamental problem remains the same: it is impractical in a single numerical simulation to resolve both the gravitational wave scale and all the small scales to which the physics transfers energy and contains sizeable dynamics. The job of \emph{turbulence modelling} is to replace the idealized model (that may be correct on all scales) with a practical model (that may be incorrect on very small scales, but captures the correct physics when simulated on practical scales).

In this review we will first discuss, in \sect~\ref{sec:ns_conditions}, the properties of neutron star matter and the expected impact on turbulence in mergers. We will then outline the mathematical background for constructing models for turbulent fluids in \sect~\ref{sec:modelling}, before looking at current results in relativistic fluids in \sect~\ref{sec:grles}. Finally we will outline some potential future directions for the field in \sect~\ref{sec:future}.

\section{Turbulence in hot neutron star matter}\label{sec:ns_conditions}

\begin{figure}
    \centering
    \includegraphics[width=0.9\textwidth]{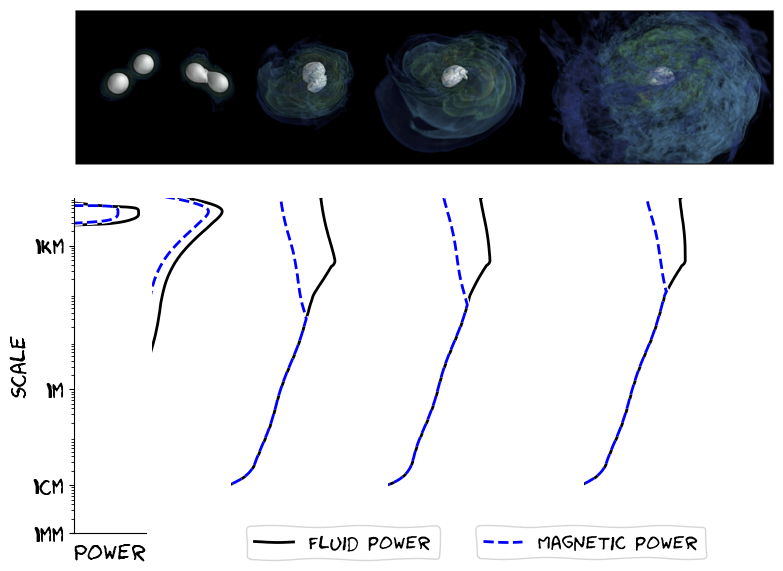}
    \caption{A sketch of the different timescales expected in a neutron star merger. In the inspiral phase both the fluid and magnetic power is contained nearly completely in long scales (small wavenumbers $\gls{k-wave} \propto 1/\gls{lengthscale}$). At merger both are disrupted. Shearing instabilities push power into all fluid lengthscales. This rapidly leads to fully developed Kolmogorov turbulence with the power following a $\gls{lengthscale}^{5/3}$ or $\gls{k-wave}^{-5/3}$ spectrum. The magnetic field power is rapidly disrupted, with the long scale power following a Kazantsev $\gls{lengthscale}^{-3/2}$ or $\gls{k-wave}^{3/2}$ spectrum. As the evolution progresses the cascade will develop until both fluid and magnetic field power follows the typical $\gls{lengthscale}^{5/3}$ cascade. The cascade cuts off below the turbulent lengthscale, with the power falling off exponentially fast due to viscous damping. Physical mechanisms governing this behaviour and best approximations for the values involved are given in \sect~\ref{sec:ns_conditions}. The merger visualization is adapted from Ciolfi, \textit{Binary Neutron Star Mergers After GW170817}, Front.~Astron.~Space Sci., 30 June 2020. Reproduced with permission.}
    \label{fig:timescale_sketch}
\end{figure}

As previously anticipated, in \ac{BNS} mergers turbulence first appears in the contact region between the stars, which is \ac{KH}-unstable. The \ac{KH} instability creates vortices with typical scale of the order of the width of the shear layer, corresponding to the fastest growing mode of the instability. These vortical structures are subject to further hydrodynamical instabilities and form progressively smaller flow structures. This is the well known \emph{turbulent cascade}. Later, turbulence develops in the remnant accretion disk as a result of the \ac{MRI}. Although some authors have suggested that convection might operate in the merger remnant, neutrino-radiation simulations appear to exclude it \citep{radice_ab-initio_2023}. However, other mechanisms, such as the Tayler–Spruit dynamo, might transport angular momentum and drive turbulence in the remnant \citep{margalit_angular-momentum_2022}. In all these regions it is expected that the end result of the turbulent cascade is the generation of a complex, chaotic flow, with structures spanning a large range of scales. The impact of this in a neutron star merger is sketched in \fig~\ref{fig:timescale_sketch}, and discussed through the rest of this section.

\subsection{General considerations}
\label{sec:ns_conditions.general}
Here, we review the basic phenomenology expected for the turbulent flows that develop in \ac{NS} mergers. The discussion will only be semi-quantitative and our aim is primarily to identify the relevant length and time scales. We will discuss the more precise mathematical formulation in \sect~\ref{sec:modelling}.

\subsubsection{Turbulent cascade}
The starting point for our discussion are the Navier-Stokes equations for Newtonian incompressible flows,
\begin{subequations}
\label{eq:navier.stokes}
\begin{align}
    \label{eq:newt_vel}
    \left( \partial_t + \gls{3vel} \cdot \nabla \right) \gls{3vel} &= -\frac{1}{\gls{rho}} \nabla \gls{pressure} + \gls{nu} \nabla^2 \gls{3vel}, \\
    \label{eq:newt_constraint}
    \nabla\cdot\gls{3vel} &= 0,
\end{align}
\end{subequations}
where \gls{3vel} is the fluid velocity, \gls{rho} is the matter density (assumed to be constant), \gls{pressure} the pressure, and \gls{nu} is the kinematic viscosity, assumed to be constant. These equations do not apply to the relativistic compressible flows present in \ac{NS} mergers, but they are sufficient to qualitatively describe the turbulence phenomenology. Moreover, on sufficiently small scales, the matter density can be considered to be constant. Similarly, the velocity fluctuations measured in a locally flat frame comoving with a sufficiently small patch of the flow can be expected to be nonrelativistic and subsonic. In such a local frame, Eq.~\eqref{eq:navier.stokes} provides a good description of the dynamics as long as magnetic stresses can be neglected, as is the case in the early stages of development of the \ac{KH} instability. We address the impact of magnetic fields in \sect~\ref{sec:ns_conditions.general.mhd}. 

From the Navier-Stokes equations we can obtain an evolution equation for the specific kinetic energy. To do so, we multiply \eqn~\eqref{eq:newt_vel}
by $\gls{3vel}$ to obtain
\begin{equation}
    \frac{1}{2} \big[ \partial_t \gls{3velmag}^2 + (\gls{3vel} \cdot \nabla) \gls{3velmag}^2 \big] = - \frac{1}{\gls{rho}} \nabla \gls{pressure} \cdot \gls{3vel} + \gls{nu} \nabla^2 \gls{3vel} \cdot \gls{3vel},
\end{equation}
where \gls{3velmag} is the magnitude of the velocity $\gls{3vel}$.
Integrating the last equation over a volume \gls{Omega},
contained in the local fluid patch in which \eqns~\eqref{eq:navier.stokes} can be considered to approximately hold, we obtain
\begin{equation}\label{eq:energy.eq}
    \frac{1}{2} \int_{\gls{Omega}} \big[ \partial_t \gls{3velmag}^2 + (\gls{3vel} \cdot \nabla) \gls{3velmag}^2 \big] \dd \gls{Omega} =
    - \frac{1}{\gls{rho}} \int_{\gls{Omega}} \nabla \gls{pressure} \cdot \gls{3vel}\, \dd \gls{Omega} + \int_{\gls{Omega}} \gls{nu} \nabla^2 \gls{3vel} \cdot \gls{3vel}\, \dd \gls{Omega} \,,
\end{equation}
where we have used the assumption that $\gls{rho} \simeq \textrm{const}$.
After integration by parts, the first integral on the RHS can be rewritten as
\begin{equation}\label{eq:energy.eq.rhs.1}
    -\int_{\gls{Omega}} \nabla \gls{pressure} \cdot \gls{3vel} \, \dd \gls{Omega} = -\int_{\partial\gls{Omega}} \gls{pressure} \gls{3vel} \cdot \gls{normal}\, \dd \Sigma + \int_{\gls{Omega}} \gls{pressure} \nabla\cdot\gls{3vel}\, \dd\gls{Omega} \,.
\end{equation}
Because turbulence is expected to be statistically isotropic and homogeneous, the first term on the RHS of \eqn~\eqref{eq:energy.eq.rhs.1} is expected to vanish. 
The second term on the RHS of \eqn~\eqref{eq:energy.eq.rhs.1} vanishes due to the incompressibility constraint.
In a similar way, we find that
\begin{equation}
   \gls{nu} \int_{\gls{Omega}} \nabla^2 \gls{3vel} \cdot \gls{3vel}\, \dd \gls{Omega} = \gls{nu} \int_{\partial\gls{Omega}} \gls{normal} \cdot (\nabla \gls{3vel}\, \gls{3vel})\, \dd \Sigma -
   \gls{nu} \int_{\gls{Omega}} \nabla\gls{3vel} : \nabla\gls{3vel}\, \dd\gls{Omega} \,.
\end{equation}
In the last term of the RHS we have used the notation $\mathbf{A}:\mathbf{B}$ to denote the inner product between two tensors:
\begin{equation}
    \mathbf{A}:\mathbf{B} = \sum_{i,k} A_{ik}\, B_{ik}\,.
\end{equation}
The first term on the RHS vanishes because of isotropy and homogeneity. Since $\gls{Omega}$ is arbitrary, we conclude that, for isotropic and homogeneous turbulence, the specific kinetic energy satisfies
\begin{equation}\label{eq:energy.v2}
   \partial_t \gls{3velmag}^2 +  (\gls{3vel}\cdot\nabla) \gls{3velmag}^2 =  - 2 \gls{nu} \lvert \nabla \gls{3vel} \rvert^2\,.
\end{equation}

Let \gls{lengthscale} be a characteristic spatial extent for the local patch of fluid under consideration. For example, $\gls{lengthscale}$ could be taken as the diameter of $\gls{Omega}$. Over times $t$ much smaller than the typical global dynamical timescale, but larger than the local eddy turnover time $\tau(\gls{lengthscale}) = \gls{lengthscale}/\gls{3velmag}(\gls{lengthscale})$, where $\gls{3velmag}(\gls{lengthscale})$ is the typical magnitude of the velocity fluctuations on the scale $\gls{lengthscale}$, the turbulence can be considered stationary \citep{pope_turbulent_2000}. By this, we mean that the velocity fluctuations on scales smaller than $\gls{lengthscale}$ are statistically stationary. That is, $\partial_t \gls{3velmag}^2 \simeq 0$ and
\begin{equation}\label{eq:energy.stationary}
   (\gls{3vel}\cdot\nabla) \gls{3velmag}^2 \simeq  - 2 \gls{nu} \lvert \nabla \gls{3vel} \rvert^2\,.
\end{equation}
In other words, over a time $t \simeq \tau(\gls{lengthscale})$, the energy injection from large scales, which is mediated by the nonlinear term $(\gls{3vel}\cdot\nabla) \gls{3velmag}^2$, achieves a balance with the energy dissipation due to viscosity. This fact is well established by experimental and numerical studies of turbulent flows \citep{frisch_turbulence_1995, pope_turbulent_2000}.

The magnitude of the energy injection or transport rate is estimated as
\begin{equation}\label{eq:kolmogorov.flux}
    (\gls{3vel}\cdot\nabla) \gls{3velmag}^2 \sim \frac{\gls{3velmag}^3(\gls{lengthscale})}{\gls{lengthscale}}\,.
\end{equation}
One of the key assumptions in Kolmogorov's theory of turbulence is that $\gls{3velmag}^3(\gls{lengthscale})/\gls{lengthscale} \sim \mathrm{const}$, independently of $\gls{lengthscale}$ over a large range of length scales, the so-called \emph{inertial range}. In other words, $\gls{3velmag}^3(\gls{lengthscale})/\gls{lengthscale}$ is independent of the extent of the region $\gls{Omega}$, if that falls within the inertial range. Often Kolmogorov's law is formulated in terms of the velocity spectrum
\begin{equation}
    \hat{E}(k) = \int \gls{3velmag}^2\, e^{-i\, k\, \gls{lengthscale}}\, {\rm d}\gls{lengthscale}\,.
\end{equation}
From Kolmogorov's hypothesis $\gls{3velmag}^2 \sim \gls{lengthscale}^{2/3}$. This implies that the spectrum has the famous power law scaling $\hat{E} \sim k^{-5/3}$, see, e.g.,~\citet{frisch_turbulence_1995} for a more detailed derivation.
According to Kolmogorov's phenomenology, turbulence transports kinetic energy from large scales, at which it is injected, to small scales, where it is converted into internal energy. The cascade proceeds until the scale at which the energy transfer rate is balanced by the viscous energy dissipation rate. The latter can be estimated as
\begin{equation}\label{eq:kolmogorov.diss}
    2 \gls{nu} \lvert \nabla \gls{3vel} \rvert^2 \sim \gls{nu} \left( \frac{\gls{3velmag}(\gls{lengthscale})}{\gls{lengthscale}} \right)^2\,,
\end{equation}
so the dissipation scale \gls{ld}
can be estimated by equating \eqns~\eqref{eq:kolmogorov.flux} and \eqref{eq:kolmogorov.diss}, giving
\begin{equation}\label{eq:dissipation.scale}
    \gls{ld} \sim \left( \frac{\gls{nu}}{\gls{l0} \gls{3velmag}(\gls{l0})} \right)^{3/4} \gls{l0} \,,
\end{equation}
where we have taken \gls{l0}
to be the largest scale of the inertial range for which $\gls{3velmag}^3(\gls{lengthscale})/\gls{lengthscale} \sim \mathrm{const}$. This outer scale is typically related to a global scale of the system.  For example, in the \ac{KH}-unstable layer in \ac{BNS} mergers $\gls{l0} \simeq \SI{1}{\kilo\metre}$. In the remnant accretion disk $\gls{l0} \simeq \gls{H} r$, where $\gls{H} \simeq 1/3$ is the disk thickness and $r$ is the cylindrical radius. 

The quantity in parenthesis in \eqn~\eqref{eq:dissipation.scale} is the inverse of the Reynolds number \gls{Re},
\begin{equation}
    \gls{Re} = \frac{\gls{l0} \gls{3velmag}(\gls{l0})}{\gls{nu}}\,,
\end{equation}
which measures the relative importance of inertial and viscous terms in the Navier-Stokes equations. 
Experimental studies of transitional flows, where the balance between inertial and viscous effects changes, show that they typically become turbulent when $\gls{Re} \gtrsim \num{1000}$ \citep{frisch_turbulence_1995}, although the value of the critical Reynolds number for transition to turbulence is not universal. This means that, for turbulent flows, $\gls{ld} < \num{0.005}\, \gls{l0}$ and the ratio $\gls{l0}/\gls{ld}$ grows as $\gls{Re}^{3/4}$. Such a large range of scales significantly limits simulations that resolve all scales of the system, the so called \ac{DNS} calculations, to flows with modest $\gls{Re}$ numbers. Indeed, the cost of \ac{DNS} calculations scales as $\gls{Re}^3$, when using explicit time integration schemes, such as those employed in numerical relativity. As we will see shortly, the Reynolds numbers encountered in \ac{NS} mergers are extremely large, so \ac{DNS} simulations are unfeasible for the foreseeable future.

\subsubsection{Magnetohydrodynamical effects}
\label{sec:ns_conditions.general.mhd}

In the presence of magnetic fields, the Navier-Stokes equations are modified due to the appearance of the Lorentz force. In CGS units, the momentum equation becomes
\begin{equation}
    \label{eq:navier.stokes.mhd}
    \left( \partial_t + \gls{3vel}\cdot\nabla \right) \gls{3vel} = \frac{1}{4\pi \gls{rho}}(\nabla\times\gls{B})\times\gls{B} - \frac{1}{\gls{rho}} \nabla \gls{pressure} + \gls{nu} \nabla^2 \gls{3vel}\,,
\end{equation}
where \gls{B} is the magnetic field. The velocity can still be taken as incompressible, $\nabla\cdot\gls{3vel} = 0$. The magnetic field is solenoidal and evolves according to the induction equation:
\begin{equation}\label{eq:induction}
    \partial_t \gls{B} + \nabla\times (\gls{B}\times\gls{3vel}) = \gls{eta} \nabla^2 \gls{B}\,,\quad \nabla\cdot\gls{B} = 0\,,
\end{equation}
where \gls{eta} is the magnetic resistivity. \eqns~\eqref{eq:navier.stokes.mhd} and \eqref{eq:induction} are the equations of classical, incompressible \ac{MHD}. Once again, the classical description is not adequate to describe the dynamics of general-relativistic flows. However, these equations can be used to describe the small-scale flow dynamics in \ac{NS} merger.

The Lorentz force can be decomposed as
\begin{equation}
     \frac{1}{4\pi} (\nabla\times\gls{B})\times\gls{B} = \frac{1}{4\pi} (\gls{B}\cdot\nabla)\gls{B} - \frac{1}{8\pi} \nabla \gls{Bmag}^2\,.
\end{equation}
The first term on the RHS corresponds to the so-called magnetic tension force, while the second term describes the magnetic pressure. This decomposition motivates the introduction of the plasma \gls{beta} parameter
\begin{equation}
    \gls{beta} = \frac{\gls{pressure}}{\gls{Bmag}^2/8\pi}\,,
\end{equation}
which measures the relative contribution of hydrodynamic and magnetic forces. In the bulk of a \ac{BNS} merger remnant, at typical densities ${\sim}\SI{5e14}{\gram\per\centi\metre\cubed}$, the pressure arising from repulsive interaction between the nucleons is ${\sim} \SI{e34}{\dyn\per\centi\metre\squared}$ and increases rapidly with density~\citep{hebeler_equation_2013}. This implies that equipartition between the magnetic and fluid pressure in the bulk of \ac{BNS} merger remnants would require fields strengths significantly in excess of $\SI{e17}{\Gauss}$. That is more than two orders of magnitude larger than that in the most magnetized \ac{NS} known~\citep{olausen_mcgill_2014}. Even in the accretion disk formed in \ac{NS} merger remnants, where the typical density is ``only''  ${\sim}\SI{e12}{\gram\per\centi\metre\cubed}$, the pressure is ${\sim} \SI{e30}{\dyn\per\centi\metre\squared}$, implying that magnetic pressure forces can be neglected for fields below $\SI{e15}{\Gauss}$. Indeed, the plasma $\gls{beta}$ values found in \ac{NS} merger remnants in \ac{GRMHD} simulations are typically larger than ${\sim} 100$ \citep{kiuchi_high_2015,palenzuela_turbulent_2022}, with the exception of magnetically dominated regions in the disk corona and in the jet.

The previous discussion would seem to suggest that magnetic stresses can be neglected when considering the cascade of energy to small scales. However, this ignores two important facts. First, magnetic stresses are not isotropic, so they can transport momentum (and angular momentum) in a shear flow. Such transverse momentum exchange is absent for an ideal fluid, so even a small magnetic field can change the dynamics in a qualitative way. Second, there is an important difference between inertial ``forces'', which dominate the turbulent cascade, and magnetic stresses. At a given scale $\gls{lengthscale}$, it is always possible to cancel the contribution of the larger scale flow in the $(\gls{3vel}\cdot\nabla)\gls{3vel}$ term of the momentum equation by using a frame in which the mean velocity vanishes. In other words, inertial forces depend only on the velocity fluctuations at a given scale $\gls{3vel}(\gls{lengthscale})$ and not on the absolute velocity. The Lorentz force, instead, is frame invariant, so it cannot be cancelled with a frame transformation \citep{beresnyak_mhd_2019}. 

In the absence of viscosity or resistivity, the hydrodynamic turbulent cascade proceeds until a scale $\gls{lB}$ at which magnetic tension and inertial forces balance:
\begin{equation}
    \frac{\gls{Bmag}^2}{4\pi} \sim \gls{rho}\, \gls{3velmag}^2(\gls{lB}) \sim \gls{rho}\, \left( \frac{\gls{lB}}{\gls{l0}} \right)^{2/3} \gls{v0}^2 \implies \gls{lB} \sim \frac{\gls{l0}}{4\pi\gls{rho}^{3/2}} \frac{\gls{Bmag}^3}{\gls{v0}^3}\,.
\end{equation}
At smaller scales the turbulent cascade changes character. Magnetic tension inhibits the motion of the flow in the plane perpendicular to the magnetic field direction and turbulence becomes strongly anisotropic. \ac{MHD} turbulence also exhibit an inverse cascade that can create large-scale magnetic structures. A detailed discussion of \ac{MHD} turbulence is beyond the scope of this review. We refer the interested reader to \citet{beresnyak_mhd_2019, schekochihin_mhd_2021} for a discussion of the current state of the field. See also~\citet{biskamp_magnetohydrodynamic_2003, kulsrud_plasma_2004} for textbook introductions.

\subsection{Turbulence in RMNSs}\label{sec:general.rmns}
For the typical thermodynamic conditions of matter in \ac{NS} mergers, momentum transfer due to neutrino diffusion provides the dominant source of viscosity. The kinematic neutrino viscosity \gls{nu_neutrino} can be estimated as \citep{van_den_horn_transport_1984,guilet_magnetorotational_2017}
\begin{equation}
    \gls{nu_neutrino} \simeq \frac{4\, \epsilon_\nu\, \langle \gls{lambdanu} \rangle}{15\, \gls{rho}\, c}\,,
\end{equation}
where $\epsilon_\nu$ is the average neutrino energy and $\langle \lambda_\nu \rangle$ is an energy averaged scattering mean free path for neutrinos. In the conditions relevant for the RMNS this yields \citep{thompson_neutron_1993}
\begin{equation}
    \gls{nu_neutrino} \simeq \num{2e8}\, f(\gls{Yp})\, \left(\frac{\gls{T}}{\SI{30}{\mega\electronvolt}}\right)\, \left(\frac{\gls{rho}}{\SI{5e14}{\gram\per\centi\metre\cubed}} \right)^{-4/3}\ \si{\centi\metre\squared\per\second}\,,
\end{equation}
where \gls{T} is the temperature,
$f(\gls{Yp}) = \big[\gls{Yp}^{1/3} + (1 - \gls{Yp})^{1/3}\big]^{-1}$ and $\gls{Yp} = \gls{np}/(\gls{np} + \gls{nn}) \simeq \num{0.05}$ is the proton fraction, with $\gls{np},\gls{nn}$ being the proton and neutron numbers. In the case of the \ac{KH} unstable interface between the \acp{NS}, a typical velocity is given by the relative velocity between the two stars, $\gls{v0} \simeq \SI{0.2}{\c}$. A typical length scale is given by the width of the shear layer, $\gls{l0} \simeq \SI{1}{\kilo\metre}$, so the Reynolds number can be estimated as
\begin{equation}
    \gls{Re} = \frac{\gls{l0} \gls{v0}}{\gls{nu_neutrino}} \simeq \num{4e6}\,.
\end{equation}
Accordingly the dissipation scale can be estimated to be
\begin{equation}\label{eq:reynolds.kh.naive}
    \gls{lnu} \sim \gls{Re}^{-3/4}\, \gls{l0} \simeq \SI{1.1}{\centi\metre}\,.
\end{equation}
However, neutrinos act as an effective source of viscosity only on scales $\gls{lengthscale} \gg \gls{lambdanu}$, the latter being the neutrino mean free path. The neutrino mean free path can be estimated as \citep{thompson_neutron_1993}
\begin{equation}\label{eq:lambda.nu}
    \gls{lambdanu} \simeq \num{100}\, f(\gls{Yp}) \left(\frac{\gls{T}}{\SI{30}{\mega\electronvolt}}\right)^{-3}\, \left(\frac{\gls{rho}}{\SI{5e14}{\gram\per\centi\metre\cubed}} \right)^{-1/3}\, \si{\centi\metre}.
\end{equation}
Since $\gls{lambdanu} \gg \gls{lnu}$, neutrino viscosity is not effective at damping the turbulent fluctuations, which continue to cascade to smaller scales. 

The dominant source of viscosity on scales smaller than $\gls{lambdanu}$ is electron scattering, which induces an effective kinematic viscosity\footnote{Here we are assuming equal number of protons and electrons.} \citep{thompson_neutron_1993}
\begin{equation}
    \gls{nu_e} \simeq 2\, \gls{Yp}\ \si{\centi\metre\squared\per\second}.
\end{equation}
The Reynolds number of the flow is thus much larger than that implied by \eqn~\eqref{eq:reynolds.kh.naive}. A better estimate is
\begin{equation}
    \gls{Re} = \frac{\gls{l0} \gls{v0}}{\gls{nu_e}} \simeq \num{5e15}.
\end{equation}
The corresponding dissipation scale $\gls{ld}$ is of the order of one nanometer.

The scale at which \ac{MHD} effects become dominant is
\begin{equation}
    \gls{lB} \simeq \SI{3.3}{\centi\metre}\ \left( \frac{\gls{rho}}{\SI{5e14}{\gram\per\centi\metre\cubed}} \right)^{-3/2}
    \left( \frac{\gls{Bmag}}{\SI{e16}{\Gauss}}\right)^{3} \left( \frac{\gls{v0}}{\SI{0.2}{\c}}\right)^{-3}\,.
\end{equation}
This implies that the back-reaction of the magnetic field becomes dominant at scales that are much larger than the dissipation length scale.

\subsection{Turbulence in the accretion disk}

A similar situation is found in the remnant accretion disk. The inner portion of the accretion disk has densities of the order of $\SI{e12}{\gram\per\centi\metre\cubed}$ and temperatures around $\SI{10}{\mega\electronvolt}$. In such conditions, nucleons are not degenerate and the neutrino viscosity can be estimated as \citep{keil_ledoux_1996}
\begin{equation}
    \gls{nu_neutrino} = \num{5.6e10} \left(\frac{\gls{T}}{\SI{10}{\mega\electronvolt}}\right)\, \left(\frac{\gls{rho}}{\SI{e12}{\gram\per\centi\metre\squared}} \right)^{-4/3}\ \si{\centi\metre\squared\per\second}.
\end{equation}
The typical scale of the turbulent motion in the disk is $\gls{l0} = \gls{H} r$, where $\gls{H} \simeq 1/3$ is the disk thickness and $r \simeq \SI{30}{\km}$ is the typical orbital radius of the densest part of the disk. Using the orbital velocity
\begin{equation}
    \gls{v0} \simeq \sqrt{\frac{G M}{r}} \simeq \num{0.4} \left(\frac{M}{3\ M_\odot}\right)^{1/2}\, \left(\frac{r}{\SI{30}{\km}}\right)^{-1/2}\, c\,,
\end{equation}
we find $\gls{Re} \simeq \num{e5}$. However, as was the case in the \ac{KH}-unstable region in the \ac{BNS} context discussed in \sect~\ref{sec:general.rmns}, the corresponding viscous scale has $\gls{lnu} \ll \gls{lambdanu}$. The latter can be estimated using \eqn~\eqref{eq:lambda.nu} to be $\gls{lambdanu} \simeq \SI{165}{\metre}$. On scales smaller than $\gls{lambdanu}$ the turbulent cascade continues all the way to the electron viscosity scale $\gls{le} \simeq \SI{5}{\nm}$. The associated Reynolds number is ${\sim} \num{2e16}$. If the disk is magnetized, as it is likely to be, then magnetic stresses become dominant at scales smaller than
\begin{equation}
    \gls{lB} \simeq \SI{50}{\centi\metre} \left( \frac{\gls{rho}}{\SI{e12}{\gram\per\centi\metre\cubed}} \right)^{-3/2}
    \left( \frac{\gls{Bmag}}{\SI{e15}{\Gauss}}\right)^{3} \left( \frac{\gls{v0}}{\SI{0.4}{\c}}\right)^{-3}\,.
\end{equation}
As is the case for massive neutron star remnants, magnetic stresses are likely to substantially alter the hydrodynamic turbulent cascade well before the dissipation range has been reached.

\subsection{Open problems}

There are two main ways in which turbulence impacts the postmerger evolution of binary mergers involving \acp{NS}.
First, turbulence drives the redistribution of the angular momentum in the remnant, driving accretion and mass ejection. Second, turbulence might strongly amplify the initial magnetic field, which, in turn, can drive relativistic outflows and power jets from these systems. 

In the case of \ac{BNS} mergers, it is known that differential rotation is necessary for the stability of the \ac{RMNS}. However, the impact of angular momentum transport on the \ac{RMNS} is an open question. As we discuss in \sect~\ref{sec:grles}, depending on the timescale over which the inner core of the \ac{RMNS} reaches solid body rotation, angular momentum transport could accelerate, or significantly delay, and even possibly avert, the collapse of the remnant. Understanding the role of turbulence on the life time of the remnant is necessary to interpret any constraints on the lifetime of \acp{RMNS}.

Another open question concerns the topology of the magnetic fields amplified by the \ac{KH} and the \ac{MRI} in these systems. Large-scale fields, with characteristic field lines whose radius of curvature is comparable to the size of the system, could  power ultrarelativistic outflows. It would also determine the evolution of \acp{RMNS} that do not collapse even after having achieved solid-body rotation. There is reason to believe that the topology of the magnetic field lines after \ac{BH} formation cannot be too simple. The reason is that \ac{GRMHD} simulations of \ac{BH} accretion show that, in the presence of large scale poloidal flux, magnetic flux is accumulated onto the \ac{BH} until magnetic pressure arrests the accretion flow \citep{narayan_magnetically_2003}. This is the so-called \ac{MAD} state. \ac{MAD} accretion flows generate extremely powerful jets, with energy output $E_K \sim \si{\Mass\disk\c\squared}$. In the \ac{NS} merger case, $\si{\Mass\disk} = \SI{0.1}{\Mass\Sun}$, so the resulting jets would have a total kinetic energy of up to ${\sim} \SI{e53}{\erg}$, which is much larger than the  typical (${\sim} \SI{e50}{\erg}$) total energy released by \ac{SGRB} engines~\citep{fong_decade_2015, mooley_superluminal_2018, ghirlanda_compact_2019}. Tangled fields, or fields with intermittent polarity, would more naturally explain the energetics of \acp{SGRB}.

\section{Turbulence Modelling}\label{sec:modelling}

The previous section shows that ideal (magneto) hydrodynamics may not be the correct theory on all scales, but is likely to be a reasonable theory on scales that can be resolved in numerical simulations. However, the physical transfer of energy-momentum from resolved scales to shorter scales will mean the ideal model will become steadily less accurate as turbulent effects become important. The modelling challenge becomes that of finding an effective theory that captures the impact of the short-scale physics using only information available at the scales we resolve. This may sound impossible, as the information needed is lost: however, Newtonian approaches suggest a range of possibilities.

\subsection{Newtonian Reynolds equations}

To motivate the modelling in relativity, we start with the standard Newtonian approach, following \citet{pope_turbulent_2000}. Assume the equations of hydrodynamics hold at all scales of interest. Assume that we can resolve some, but not all, of the scales (for reasons of complexity, or numerical resolution, for example). We then want to derive the effective equations of motion at scales that we can resolve.

There are two essential viewpoints. One is that the behaviour on the finest possible scales is essentially unknown, or random. In this case we treat the ``true'' fine scale behaviour as stochastic, where the flow variables are truly random variables. We then get a mean-field theory on the resolved scales by averaging over all possible realisations of the flow variables.

The second viewpoint is that we know the behaviour on the finest possible scales. In this case, resolving some of the scales means averaging over space and/or time to get a mean-field theory that holds on the resolved scales. This has a number of subtle issues in relativity and so will be looked at later.

To illustrate how this works in the Newtonian case, consider again the constant density incompressible Navier-Stokes equations\footnote{We should warn the reader that the pressure has a different character and role in the incompressible and compressible cases. In the compressible case the pressure is given by a closure relation -- the \ac{EOS} -- which includes thermodynamical and microphysical aspects of the matter. In the incompressible case the pressure is a Lagrange multiplier ensuring that the constraint~\eqref{eq:newt_constraint} is satisfied. As our motivating example of neutron star mergers requires the compressible case we shall develop the theory as if the pressure is given by an \ac{EOS}. Note that some steps and approaches in the standard turbulence literature are not useful in our case, as they rely on incompressibility to reformulate the pressure terms.}
\begin{subequations}
\begin{align}
    \tag{\ref{eq:newt_vel}}
    \left( \partial_t + \gls{3vel} \cdot \nabla \right) \gls{3vel} &= -\frac{1}{\gls{rho}} \nabla \gls{pressure} + \gls{nu} \nabla^2 \gls{3vel}, \\
    \tag{\ref{eq:newt_constraint}}
    \nabla\cdot\gls{3vel} &= 0.
\end{align}
\end{subequations}
We now want to construct an averaging or filtering operation for the key fields, such as the velocity \gls{3vel} which under the operation gives $\ave{\gls{3vel}}$.

\subsubsection{Statistical averaging}

Assume the probability distribution function of the velocity field is $f$, which depends on space and time. Write the statistical averaging as
\begin{equation}
    \ave{{\gls{3vel}}(\bm{x}, t)} = \int {\VV} f({\VV}; \bm{x}, t) \, \dd[3]{\VV}.
\end{equation}
We can then write the velocity field in terms of its average and fluctuation as
\begin{equation}
    \gls{3vel} = \ave{\gls{3vel}} + \fluct{\gls{3vel}}.
\end{equation}
Importantly there is a frame in which the fluctuation averages to zero, $\ave{\gls{3vel}} = 0$. The mean field model will give us the equations of motion in terms of the averaged flow variables, which in this simplified case is the velocity field alone.

A crucial assumption to make is that the underlying fields are continuous. At sufficiently short scales this should be reasonable, as the higher-derivative terms associated with viscosity should prevent wave breaking. This assumption guarantees that statistical averaging commutes with differentiation\footnote{For a more general approach, see \citet[chapter 12]{pope_turbulent_2000}.}, as, for example,
\begin{subequations}
\begin{align}
    \ave{\pdv{v(t)}{t}} &= \ave{\lim_{\Delta t \to 0} \frac{ v(t + \Delta t) - v(t)}{\Delta t}} \\
    &= \lim_{\Delta t \to 0} \frac{ \ave{v(t + \Delta t)} - \ave{v(t)}}{\Delta t} \\
    &= \pdv{\ave{v(t)}}{t}.
\end{align}
\end{subequations}

This approach is often associated with \ac{RANS}, which is central to the development of turbulence models in the Newtonian theory, but not to relativistic applications.

\subsubsection{Filtering}
\label{sec:modelling.filtering}

The alternative approach is to assume that we are resolving the behaviour on some lengthscale $L$ and to introduce a filtering kernel $G(\bm{x}'; \bm{x})$ that acts on that lengthscale. A typical example would be the Gaussian kernel $G = (4 \pi L)^{-1/2} \exp \left\{ -\left( \| \bm{x} - \bm{x}' \| / (4 L) \right)^2 \right\}$.

The averaging operation is then given by
\begin{equation}
    \label{eq:filtering.kernel}
    \ave{\gls{3vel}}(\bm{x}) = \int \gls{3vel}(\bm{x}')\; G \left( \bm{x}'; \bm{x} \right) \, \dd[3]{\bm{x}'}.
\end{equation}

This approach is often associated with \ac{LES}, and has been the base of most approaches used in relativity.

\subsubsection{Reynolds equations}

Taking the Navier-Stokes equations, substituting in the average/fluctuation form of the velocity, and averaging, we find the \emph{Reynolds equations}
\begin{subequations}
\label{eq:newtonian_reynolds}
\begin{align}
    \label{eq:newtonian_reynolds_momentum}
    \left(\partial_t + \overline{\gls{3vel}} \cdot \nabla \right) \overline{\gls{3vel}}
    &= -\frac{1}{\gls{rho}} \nabla \ave{\gls{pressure}} + \gls{nu} \nabla^2 \ave{\gls{3vel}} - \nabla \cdot \ave{ \fluct{\gls{3vel}} \otimes \fluct{\gls{3vel}}}, \\
    \label{eq:newtonian_reynolds_continuity}
    \nabla \cdot \ave{\gls{3vel}} &= 0 = \nabla \cdot \fluct{\gls{3vel}}.
\end{align}
\end{subequations}
We see that the Reynolds equations match the Navier-Stokes equations except for the term involving the \emph{Reynolds stresses} $\ave{\fluct{\gls{3vel}} \otimes \fluct{\gls{3vel}}}$. These stresses act like an additional source of dissipation, and result from the fine-scale behaviour we are unable to resolve.

\subsection{The closure problem}

When evolving the Reynolds equations, we are assuming we have initial data and boundary conditions for the mean flow $\ave{\gls{3vel}}$, but no information on the fluctuations $\fluct{\gls{3vel}}$. Therefore \eqns~\eqref{eq:newtonian_reynolds} are not a closed system until we provide information on the fluctuations. This is the closure problem: we need to provide sufficient information on the fine scale behaviour in order to solve the coarse scale problem. As we are solving only for the coarse scale because working at all scales is impractical, this means giving some summary statistics of the correlations that lead to the Reynolds stresses $\ave{\fluct{\gls{3vel}} \otimes \fluct{\gls{3vel}}}$.

It is useful to decompose the Reynolds stresses further. A standard choice is to define the \emph{turbulent kinetic energy}
\begin{equation}
    \gls{tke} = \tfrac{1}{2} \ave{\fluct{\gls{3vel}} \cdot \fluct{\gls{3vel}}}.
\end{equation}
Note that dimensionally $\gls{tke}$ is a \emph{specific} energy, but in the incompressible case this is often not mentioned. The Reynolds stress can then be written as an isotropic and anisotropic part,
\begin{equation}
    \ave{\fluct{\glslink{3vel}{v^i}} \fluct{\glslink{3vel}{v_j}}} = \tfrac{2}{3} \gls{tke} \delta^i_j + \gls{anisotropic-stress}.
\end{equation}
We can then rewrite all stress terms in the momentum equation~\eqref{eq:newtonian_reynolds_momentum} as
\begin{equation}
    -\frac{1}{\gls{rho}} \, \nabla \ave{\gls{pressure}}  - \nabla \cdot \ave{\fluct{\gls{3vel}} \otimes \fluct{\gls{3vel}}} = -\frac{1}{\gls{rho}} \nabla \left( \ave{\gls{pressure}} + \tfrac{2}{3} \gls{rho} \, \gls{tke} \right) - \nabla \cdot \glslink{anisotropic-stress}{a}.
\end{equation}
Thus the isotropic term $\tfrac{2}{3} \gls{rho} \gls{tke}$ can be absorbed in a modified pressure. Only the anisotropic term $\glslink{anisotropic-stress}{a_{ij}}$ is effective in transporting momentum in novel ways.

This decomposition immediately allows us to write down the first solution to the closure problem. The \emph{turbulent-viscosity hypothesis} of Boussinesq assumes that the anisotropic stress $\glslink{anisotropic-stress}{a_{ij}}$ is directly proportional to the mean rate of strain of the average flow,
\begin{equation}
    \glslink{anisotropic-stress}{a_{ij}} = \gls{nu_T} \left( \partial_j \ave{\glslink{3vel}{v_i}} + \partial_i \ave{\glslink{3vel}{v_j}} \right).
\end{equation}
The \emph{turbulent viscosity} or \emph{eddy viscosity} \gls{nu_T} is a scalar coefficient. The Reynolds equations immediately become
\begin{equation}
    \left(\partial_t + \overline{\gls{3vel}} \cdot \nabla \right) \overline{\gls{3vel}} = -\frac{1}{\gls{rho}} \nabla \left( \ave{\gls{pressure}} + \tfrac{2}{3} \gls{rho} \, \gls{tke} \right) + \gls{nu_eff} \nabla^2 \ave{\gls{3vel}},
\end{equation}
where the \emph{effective viscosity} is
\begin{equation}
    \gls{nu_eff} = \gls{nu} + \gls{nu_T}.
\end{equation}
This hypothesis returns the Reynolds equations to \emph{exactly} the form of the original incompressible Navier-Stokes equations, only with an effective viscosity coefficient and a modified mean pressure.

Experiments in Newtonian flows have shown that the turbulent-viscosity hypothesis may be extremely inaccurate. Nonetheless, the intuition of introducing modified pressures and viscous coefficients to capture the unresolvable small scale physics remains important in most turbulence models.

\subsection{Relativity via the equations of motion}
\label{sec:relativity.valencia}

The steps in the previous sections can be directly applied to the relativistic Euler equations once we have the equations of motion. We illustrate the procedure following \citet{radice_binary_2020} which uses the Valencia approach. The conserved current (usually thought of as the baryon mass current) $\gls{J_current} = m \gls{n_current}$, and the stress-energy tensor
\begin{equation}
    \label{eqref:stress.energy.ideal}
    \gls{stress-energy} = \gls{e} \gls{u} \glslink{u}{u_b} + \gls{pressure} (\delta^a_b + \gls{u} \glslink{u}{u_b}),
\end{equation}
where \gls{e} is the total energy density and $\gls{u}$ the 4-velocity, are decomposed with respect to the normal to the spatial slice \gls{N_slice}. The resulting equations of motion are in balance law form
\begin{equation}
    \label{eq:valencia.balance}
    \partial_t \left( \sqrt{\gls{gamma}} \vec{q} \right) + \partial_j \left( \gls{alpha_3metric} \sqrt{\gls{gamma}} \, \vec{f}^{(j)} \right) = \vec{s}.
\end{equation}
The conserved quantities $\vec{q} = (\gls{D}, \gls{Si}, \gls{E})$ are projections of the conserved current and the stress-energy tensor along the normal to the $t=\text{const.}$ hypersurfaces \gls{N_slice}. The fluxes are
\begin{subequations}
    \label{eq:valencia}
    \begin{align}
        \label{eq:valencia.D}
        \vec{f}^{(j)}_{\gls{D}} &=  \left( \glslink{r3vel}{v^j} + \glslink{N_slice}{N^j} \right) \gls{D} , \\
        \label{eq:valencia.S}
        \vec{f}^{(j)}_{\gls{Si}} &=  \glslink{Sij}{S^j_i} + \gls{Si} \glslink{N_slice}{N^j} , \\
        \label{eq:valencia.E}
        \vec{f}^{(j)}_{\gls{E}} &=  \glslink{Si}{S^j} + \gls{E} \glslink{N_slice}{N^j} ,
    \end{align}
\end{subequations}
which contain nonlinear combinations of the conserved variables, whilst the sources are linear in the conserved variables.

It is worth noting that, by reframing the continuity equation $\nabla_a \gls{J_current} = 0$ in terms of the relativistic material derivative $m \gls{ndot} = m \gls{u} \nabla_a \gls{n}$ we have
\begin{equation}
    \gls{ndot} = - \gls{n} \nabla_a \gls{u},
\end{equation}
and so on scales where the flow is stationary in the sense that $\gls{ndot} \simeq 0$, the flow is incompressible in the sense that $\nabla_a \gls{u}$. This gives the link to the assumption of incompressibility in the Newtonian context.

The averaging operation can be directly applied to the equations of motion~\eqref{eq:valencia}, but various assumptions are needed to close the system. In~\citet{radice_binary_2020} the first assumption is that metric quantities are unaffected by averaging (as the curvature lengthscale is long compared to the fluid scales). The second assumption is needed to close the nonlinear terms in the fluxes, which can be written
\begin{subequations}
    \label{eq:valencia.closure}
    \begin{align}
        \label{eq:valencia.closure.tau}
        \ave{\gls{Sij}} &= \ave{\gls{Si}} \ave{\glslink{r3vel}{v_j}} + \ave{\gls{pressure}} \delta_{ij} + \glslink{tau_rs}{\tau_{ij}}, \\
        \label{eq:valencia.closure.mu}
        \ave{\gls{D} \glslink{r3vel}{v^i}} &= \ave{\gls{D}} \ave{\glslink{r3vel}{v^i}} + \mu^i.
    \end{align}
\end{subequations}
The closure terms are \gls{tau_rs} which is the analogue of the Reynolds stress introduced in the Newtonian case, and a new term $\mu^i$ which describes turbulent mass diffusion. The explicit closure is detailed in \sect~\ref{sec:closure.smagorinsky}. 

There are two additional nonlinearities that are less obvious and require closure relations. First, the equation of state relates the pressure $\gls{pressure}$ to the other thermodynamic potentials. After averaging, we need a closure relation when the functional form of the equation of state is applied to the averaged quantities, as
\begin{equation}
    \label{eq:valencia.closure.p}
    \ave{\gls{pressure}} = \gls{pressure}\left(\ave{\gls{D}}, \ave{\gls{Si}}, \ave{\gls{E}} \right) + \gls{bulk-viscous}.
\end{equation}
In most approaches it is assumed that $\gls{bulk-viscous} = 0$ as the turbulent fluctuations are subsonic at sufficiently small scales, so corrections to the pressure are expected to be subdominant. This term is present in the Newtonian compressible case. 

The second additional closure is in the recovery of the three-velocity $\ave{\glslink{r3vel}{v^i}}$ from the coarse-grained quantities. This is a purely relativistic effect, which is linked to the ``correct'' way to average the Lorentz factor, and has been discussed in detail by~\citet{carrasco_gradient_2020,vigano_general_2020,duez_comparison_2020,radice_binary_2020}. The practical approaches suggested there all ``work'' but explicitly break the underlying 4-covariance of the theory. To retain this underlying symmetry requires a different approach.

\subsection{Relativity via an observer}
\label{sec:relativity.observer}

The previous sections essentially introduced three separate steps. First, an operation (averaging) was introduced that takes fields that vary on all scales to fields that vary only on long scales. Second, the operation is applied to the ``true'' equations of motion to create the effective field theory. Finally, additional closure relations are provided to complete the effective field theory.

Covariance can be broken in each of these three steps. In averaging the equations of motion produced by the Valencia approach in \sect~\ref{sec:relativity.valencia} the normal to the spatial slice \gls{N_slice} appears implicitly or explicitly in each step, and leads to particularly problems with (for example) the averaging of the Lorentz factor. An alternative viewpoint is that averaging applied to a single slice inherits the dependencies of the slice.

Discussions of how covariance can be retained are given in~\citet{eyink_cascades_2018} and~\citet{celora_covariant_2021}. The approach of \citet{celora_covariant_2021} will be outlined here. Fundamentally it relies on introducing an observer via their 4-velocity vector \gls{observer} and constructing the averaging operation in a (Fermi-transported) frame orthogonal to that observer. With respect to that observer it can be shown (rather than assumed) that averaging does not affect metric quantities and also commutes with covariant derivatives. The conserved currents and stress-energy tensor can also be decomposed with respect to the observer. Using notation analogous to \sect~\ref{sec:relativity.valencia}, particularly \eqn~\eqref{eq:valencia.closure}, we will find
\begin{subequations}
    \label{eq:relativity.observer.eom}
    \begin{align}
        \nabla_a \left( m \ave{\gls{n}} \gls{observer} \right) &= \nabla_a \mu^a,\\
        \nabla_a T^a_{b \text{, ideal}} &= -\nabla_a \gls{tau_rs}.
    \end{align}
\end{subequations}
If we can find a ``physical'' way of constructing the observer 4-velocity (that is, one independent of the normal to the $3+1$ slices \gls{N_slice}) then the Valencia approach can be used in the standard way with this new set of equations of motion: a tetrad is introduced with timelike leg along \gls{N_slice}, and equations~\eqref{eq:relativity.observer.eom} are projected with respect to the tetrad. As the total stress-energy tensor expressed with respect to the observer \gls{observer} is full (not diagonal, as the perfect-fluid stress-energy expressed with respect to the fine-grained 4-velocity is), the resulting equations of motion will look like those of a relativistic non-ideal fluid.

The result should be covariant, but has the equivalent complexities and problems of \emph{non-ideal} relativistic flows. The closure terms $\mu^a$ and \gls{tau_rs} are now linked to the particle drift and the non-ideal bulk, shear, and heat-transport terms, all with respect to the observer. 

\subsubsection{Thermodynamics}

When decomposing the stress-energy we will have
\begin{align}
    \gls{stress-energy} &= \tilde{\gls{e}} \gls{observer} \glslink{observer}{U_b} + \tilde{\gls{pressure}} \left( \delta^a_b + \gls{observer} \glslink{observer}{U_b} \right) + \gls{tau_rs} \\
    &= T^a_{b \text{, ideal}} + \gls{tau_rs}.
\end{align}
The terms $\tilde{\gls{e}}, \tilde{\gls{pressure}}$ come purely from projecting the stress-energy tensor with respect to this new 4-velocity. The additional $\gls{tau_rs}$ terms can be interpreted as \emph{non-ideal} terms, illustrating that stresses in the ``true'' fluid are, generically, not isotropic with respect to the mean flow.

We again see the subtlety in the standard closure relations -- the equation of state. The pressure at the fine scale is given by the \ac{EOS} in terms of the thermodynamical potentials at the fine scale: for example, $\gls{pressure} = \gls{pressure}(\gls{n}, \gls{e})$. At the averaged scale we might expect $\ave{\gls{pressure}} = \ave{\gls{pressure}} (\ave{\gls{n}}, \ave{\gls{e}})$ or $\ave{\gls{pressure}} = \ave{\gls{pressure}} (\ave{\gls{n}}, \tilde{\gls{e}})$. However, there is no guarantee that the averaging process preserves the functional form of the nonlinear closure that is the equation of state. Any differences could be formally included within a bulk viscous pressure $\gls{bulk-viscous}$. However, this highlights that we need to properly account for the thermodynamics under any averaging procedure, in addition to statistical or spacetime filtering effects.

The averaging procedure identifies, at the coarse-grained level, a number density (say $\ave{\gls{n}}$) and an energy density $\tilde{\gls{e}}$ of the mean flow. It also identifies a \emph{total} pressure $\tilde{\gls{pressure}}$. However, we are interpreting the (full) stress-energy tensor as a non-ideal fluid. That means the total pressure comprises a piece in thermodynamic equilibrium plus the bulk viscous correction.

We are free, at some level, to choose the mean \ac{EOS} (that gives the equilibrium pressure) as we want. We clearly want to relate this as closely as possible to the ``true'' (micro-scale) \ac{EOS}, but the coupling between different fluctuations means there will always be some difference. As an illustration of the differences that can arise when averaging, let us assume that the microscale \ac{EOS} is barotropic, $\gls{e} = \gls{e}(\gls{n})$, and consider the Gibbs relation
\begin{equation}
    \gls{pressure} + \gls{e} = \gls{n} \gls{chemical-potential},
\end{equation}
where the chemical potential is $\gls{chemical-potential} = \dv{\gls{e}}{\gls{n}}$. Applying the split into mean and fluctuating pieces we immediately find
\begin{subequations}
\begin{align}
    \ave{\gls{pressure}} + \ave{\gls{e}} &= \ave{\gls{n}} \ave{\gls{chemical-potential}} + \ave{\fluct{\gls{n}}\fluct{\gls{chemical-potential}}} \\
    &= \ave{\gls{n}} \gls{chemical-potential}(\ave{\gls{n}}) + \left[ \frac{\ave{\gls{n}}}{2} \gls{chemical-potential}''(\ave{\gls{n}}) + \gls{chemical-potential}'(\ave{\gls{n}}) \right] \ave{\fluct{\gls{n}}\fluct{\gls{n}}}.
\end{align}
\end{subequations}
The interpretation of this equation is that \emph{if} we use the same microscale \ac{EOS} applied to the mean flow variables (i.e., use $\gls{chemical-potential}(\ave{\gls{n}})$) then the average Gibbs relation is \emph{not} barotropic. The additional term in square brackets can be interpreted as a second parameter and would usually be thought of as $\gls{s}\gls{T}$, the piece corresponding to entropy and temperature. Alternatively it can be thought of as part of the total pressure. There is no clean distinction here.

This argument can be extended to more general \acp{EOS}. We can (implicitly) specify the coarse-grained \ac{EOS} through the thermodynamic potential representing the coarse-grained entropy, $\tilde{\gls{s}} = \tilde{\gls{s}}(\tilde{\gls{n}}, \tilde{\gls{e}})$. In the same way as above, where we noted that we are free to choose the mean (coarse-grained) \ac{EOS} as we want, we are also free to choose the coarse-grained entropy as we want. From this we can \emph{define} the coarse-grained temperature and chemical potential to obey the standard definitions
\begin{subequations}
\begin{align}
    \frac{1}{\tilde{\gls{T}}} &= \left( \pdv{\tilde{\gls{s}}}{\tilde{\gls{e}}} \right)_{\tilde{\gls{n}}} \left( \tilde{\gls{n}}, \tilde{\gls{e}} \right), \\
    - \frac{\tilde{\gls{chemical-potential}}}{\tilde{\gls{T}}} &= \left( \pdv{\tilde{\gls{s}}}{\tilde{\gls{n}}} \right)_{\tilde{\gls{e}}} \left( \tilde{\gls{n}}, \tilde{\gls{e}} \right).
\end{align}
\end{subequations}
The mean Gibbs relation is then
\begin{equation}
    \ave{\gls{pressure}} + \tilde{\gls{e}} = \tilde{\gls{n}} \tilde{\gls{chemical-potential}} + \tilde{\gls{s}} \tilde{\gls{T}} + \gls{residual-p},
\end{equation}
where \gls{residual-p} holds the residuals (the differences between the mean flow variables and their averages). At this point we have constructed (by assumption) the entropy of the mean flow, so the natural interpretation of \gls{residual-p} is as the bulk viscous pressure \gls{bulk-viscous}. It should be clear, however, that there is a direct link between the value of \gls{residual-p} and the assumed structure of the entropy of the mean flow.

\subsection{Numerics}
\label{sec:numerics}

We have seen how (statistical) averaging or (spacetime) filtering leads to equations of motion (in the Newtonian case) that contain dissipative terms. In the relativistic compressible case a similar analysis performed at the level of the stress-energy tensor leads to a full, non-ideal ``fluid'' stress tensor after averaging or filtering, even when the theory at the fine scale is purely ideal. To consider how practical these averaged or filtered theories are, we should look at how non-ideal relativistic theories are numerically evolved.

The starting point is the mean stress tensor $\ave{\gls{stress-energy}}$ and any mean charge currents, $\ave{\gls{n_current}}$ under consideration. Using the standard identity $\nabla_a V^a = (-\gls{g})^{-1/2} \partial_a \left( (-\gls{g})^{1/2} V^a \right)$, the equation of motion for the charge currents will follow as
\begin{equation}
    \nabla_a \ave{\gls{n_current}} = \gls{creation_rate} \quad \implies \quad \partial_a \left( (-\gls{g})^{1/2} \ave{\gls{n_current}} \right) = (-\gls{g})^{1/2} \gls{creation_rate}.
\end{equation}
Here $\gls{creation_rate}$ is the net creation rate of the charged species, which may arise purely from the averaging or filtering operation.

For the mean stress tensor, the Valencia approach can be used. A tetrad $\{ \gls{tetrad} \}$ is introduced with $e^b_{(0)} = \ave{\glslink{N_slice}{N^b}}$ timelike. The contractions $V^a_{(j)} = \ave{\gls{stress-energy}} \gls{tetrad}$ are therefore vectors, and stress-energy conservation and the above identity imply
\begin{equation}
    \partial_a \left( (-\gls{g})^{1/2} \ave{\gls{stress-energy}} \gls{tetrad} \right) = - (-\gls{g})^{1/2} \ave{\gls{stress-energy}} \nabla_a \gls{tetrad}.
\end{equation}
This is a \emph{balance law}, evolving the \emph{conserved variables} $(-\gls{g})^{1/2} \ave{\glslink{stress-energy}{T^0_b}} \gls{tetrad}$ in terms of their associated \emph{fluxes} $(-\gls{g})^{1/2} \ave{\glslink{stress-energy}{T^i_b}} \gls{tetrad}$ and the \emph{geometric source terms} which, involving derivatives only of the chosen tetrad, are algebraic in terms of the matter quantities.

Whilst this will give the equations of motion for a generic filtering operation as those for a non-ideal system, this is not the most practical approach for implementation. As illustrated above, the filtering operation splits the stress-energy into a piece that appears to be ideal with respect to the mean flow, and a non-ideal piece,
\begin{equation}
  \ave{\gls{stress-energy}} = T^{\text{ideal, }a}_b + \gls{tau_rs}.
\end{equation}
Again, $\gls{tau_rs}$ is the analogue of the Reynolds stress.
This means the equations of motion become
\begin{equation}
    \begin{split}
    \partial_a \left( (-\gls{g})^{1/2} T^{\text{ideal, }a}_b \gls{tetrad} \right) = - (-\gls{g})^{1/2} T^{\text{ideal, }a}_b \nabla_a \gls{tetrad} - \\ \underbrace{\left[ \partial_a \left( (-\gls{g})^{1/2} \gls{tau_rs} \gls{tetrad} \right) + (-\gls{g})^{1/2} \gls{tau_rs} \nabla_a \gls{tetrad} \right]}_{\mathcal{S}}.
    \end{split}
\end{equation}
The ``ideal'' part of the equations is implemented in a range of standard codes across the community. The additional terms $\mathcal{S}$ are those that need to be provided by the turbulent closure. Therefore, it is standard to extend and re-interpret an existing ideal fluid code, rather than implementing a full turbulence model from scratch. Care should be taken when interpreting the results, particularly on the thermodynamics.

There remains a cautionary point to note with the numerical implementation. The numerical error can, via the \emph{modified equation approach} (Sec.~\ref{sec:closure.iles} and, \citep[e.g.,][]{warming_modified_1974}), be interpreted as the numerical scheme exactly solving a modified set of equations of motion. This means there is an ambiguity in the ``non-ideal'' behaviour observed in simulation results between effects driven by numerical discretisation error and effects driven by the closure relations attempting to capture short-scale physics. Standard grid convergence tests also need careful interpretation, as the closure terms may sensibly vary depending on the resolved lengthscale as well.

\subsection{Closure relations}

There are three key approaches to the closure problem for relativistic flows that need considering.

\subsubsection{Implicit LES}
\label{sec:closure.iles}
The simplest possible turbulence closure is to neglect the turbulent stresses alltogether, i.e., setting $\tau_{ab}=0$. In this case, one relies on the intrinsic numerical dissipation of \ac{HRSC} schemes to model the effects of unresolved turbulence fluctuations. This approach, called \ac{ILES}, might appear to be unlikely to work at first. Nevertheless, \ac{ILES} has been tremendously successful in many areas of science and engineering \citep{grinstein_implicit_2007}. The theoretical foundation of this method relies on the  modified equation approach mentioned in Sec.~\ref{sec:numerics}, which we briefly sketch below. We refer to \citet{grinstein_implicit_2007} for a more systematic discussion. Consider a second-order accurate \ac{HRSC} discretization of a conservation law
\begin{equation}\label{eq:claw.generic}
    \partial_t u_a + \partial_i f^i_a(u) = 0\,.
\end{equation}
Here, second order accurate means that the numerical solution $u_h = u + \mathcal{O}(h^2)$, where $h$ is some measure of the grid spacing\footnote{For simplicity, we are ignoring error terms coming from the discretization in time.}. It is possible to show that $u_h$ is a \emph{third-order accurate} approximation to the solution of the modified equation
\begin{equation}\label{eq:claw.modified}
    \partial_t v_a + \partial_i f^i_a(v) = \partial_b {\hat\tau}^b_a(v,h)\,,
\end{equation}
where $\hat\tau^b_a(v,h)$ is a nonlinear function of $v$ which depends on the details of the numerical scheme. In other words, $u_h = v + \mathcal{O}(h^3)$ and $\hat\tau^b_a(v,h)$, sometimes called the ``numerical viscosity'', can be thought of as representing the leading order contribution to the numerical error. The \ac{ILES} methodology is based on the observation that the $\hat\tau_b^a$ of widely used numerical schemes, such as the piecewise parabolic method (PPM; \citealt{colella_piecewise_1984}), are similar to explicit closures $\tau_b^a$, such as the Smagorinsky closure described in Sec.~\ref{sec:closure.smagorinsky}. The main advantage of the \ac{ILES} method is that it is simple to implement (the only requirement is to choose the components of the numerical solver carefully). The main disadvantage is that \ac{ILES} requires that a significant fraction of the inertial range to be resolved in order to give convergent results, \citep[e.g.,][]{schmidt_numerical_2006, thornber_implicit_2007, schmidt_large_2015, radice_implicit_2015}. In practice, even though \ac{ILES} is used in most published \ac{NS} merger simulations, none of the \ac{ILES} simulations have been shown to be in a convergent regime, as discussed in more detail in Sec.~\ref{sec:grles}.

\subsubsection{Smagorinsky}
\label{sec:closure.smagorinsky}

\citet{smagorinsky_general_1963} introduced a closure model for Newtonian atmospheric flows that extends the Boussinesq hypothesis to couple the effective anisotropic stresses to the shear of the mean flow. This model has seen great success in Newtonian flows despite its simplicity. \citet{radice_general-relativistic_2017,radice_dynamics_2020} extended this to relativity, choosing the closure to be
\begin{equation}\label{eq:smagorinsky}
  \glslink{tau_rs}{\tau_{ab}} = -2 \gls{nu_T} \glslink{perp}{\perp^c_a} \glslink{perp}{\perp^d_b} (\tilde{\gls{e}} + \tilde{\gls{pressure}}) \left[ \frac{1}{2} \left( \nabla_c \glslink{vbar}{\bar{v}_d} + \nabla_d \glslink{vbar}{\bar{v}_c} \right) - \frac{1}{3} \nabla_k \glslink{vbar}{\bar{v}^k} \gls{gamma_metric} \right].
\end{equation}
Here $\glslink{perp}{\perp^c_a} = \delta^c_a + \glslink{N_slice}{N^c} \glslink{N_slice}{N_a}$ projects into the spatial slice, and $\glslink{vbar}{\bar{v}_c}$ is the 3-velocity of the mean flow in the spatial slice.

Two key features appear immediately. First, the model explicitly depends on a viscosity scalar $\nu_T$ similar to the kinematic viscosity introduced in the Newtonian case. This has to be provided based on additional physical arguments. Radice chooses, on dimensional grounds, to link $\gls{nu_T} = \gls{lmix} \gls{cs}$, coupling a lengthscale (the \emph{mixing length} \gls{lmix})
with the local speed of sound of the flow. The mixing length can then be linked to the appropriate turbulent lengthscales for the problem as discussed in \sect~\ref{sec:ns_conditions}.

Second, we see that covariance is explicitly broken as the closure relation depends on the gauge through the normal to the spatial slice \gls{N_slice}. In principle this is a problematic feature of the closure model -- detailed discussions of covariance are in~\citet{eyink_cascades_2018,celora_covariant_2021}. At present, however, all practical closure implementations share this feature.

\subsubsection{Explicit filtering}
\label{sec:closure.explicit}

In the discussion so far we have first assumed we know the short lengthscale behaviour that we then average/filter, only to discard all that knowledge as soon as we evolve the mean flow equations of motion. This aligns with the assumption that the unresolved behaviour is due to physics which we cannot access -- for example, we do not have initial data for the fluctuations.

An alternative approach is to assume that all the closure terms come from the filtering operation, and that operation is explicitly known. For example, we could argue that the numerical lengthscale is much larger than the physical dissipation lengthscales, so the filtering due to numerical discretisation dominates. This approach has been extended to relativity for successively more complex models by \citet{vigano_extension_2019,carrasco_gradient_2020,palenzuela_large_2022}.

As in \sect~\ref{sec:modelling.filtering} we assume we know the filtering kernel $G(\bm{x}'; \bm{x})$ from which we compute the filtered quantity
\begin{equation}
    \label{eq:closure.filtering.kernel}
    \ave{q}(\bm{x}) = G * q = \int q(\bm{x}')\; G \left( \bm{x}'; \bm{x} \right) \, \dd[3]{\bm{x}'}.
\end{equation}
As the Fourier transform of the convolution is a multiplication, we can write the result and its inverse in frequency space as
\begin{equation}
    \label{eq:closure.filtering.fourier}
    \mathcal{F} \left( \ave{q} \right) = \mathcal{F} \left( G \right) \mathcal{F} \left( q \right), \quad \mathcal{F} \left( q \right) = \mathcal{F} \left( G \right)^{-1} \mathcal{F} \left( \ave{q} \right).
\end{equation}
In principle this allows the original quantity to be computed given the coarse grained quantity and the filter. However, this relies on the inverse of the filter existing and being bounded, which cannot be the case when coarse graining in numerical simulations where information is lost. However, $\mathcal{F}(G)$ can be approximated by an expansion with a bounded inverse, and this used to relate filtered quantities (see~\citealt{stolz_approximate_1999} for links to approximate deconvolution methods).

In particular, using a gradient or Taylor series approximation of $\mathcal{F}(G)$ (as in~\citealt{vlaykov_nonlinear_2016,vigano_extension_2019}) it can be shown that
\begin{equation}
    \label{eq:closure.filtering.product}
    \ave{q_1 q_2} = \ave{q_1} \ave{q_2} + C \partial^k \ave{q_1} \partial_k \ave{q_2} + \mathcal{O} \left( C^2 \right)
\end{equation}
where $C$ depends on the precise form of the filtering kernel. In particular, for the Gaussian kernel $G = (4 \pi L)^{-1/2} \exp \left\{ -\left( \| \bm{x} - \bm{x}' \| / (4 L) \right)^2 \right\}$ we have that $C = L / 2$, linking the correction terms to the lengthscale of the filter.

With these results, the equations of motion in balance law form
\begin{equation}
    \tag{\ref{eq:valencia.balance}}
    \partial_t \left( \sqrt{\gls{gamma}} \vec{q} \right) + \partial_j \left( \gls{alpha_3metric} \sqrt{\gls{gamma}} \, \vec{f}^{(j)} \right) = \vec{s}
\end{equation}
can be filtered (assuming, following \citealt{vigano_extension_2019}, that the curvature lengthscale is much greater than the filtering and turbulent lengthscales, meaning the metric terms are pulled out of the filtering) as
\begin{equation}
    \label{eq:closure.filtering.eom}
    \partial_t \left( \sqrt{\gls{gamma}} \ave{\vec{q}} \right) + \partial_j  \left( \gls{alpha_3metric} \sqrt{\gls{gamma}} \, \vec{f}^{(j)} \left( \ave{\vec{q}} \right) \right) = \ave{\vec{s}} + \partial_j \glslink{tau_rs}{\vec{\tau}^{(j)}}.
\end{equation}
The analogue of the Reynolds stresses is
\begin{equation}
    \label{eq:closure.filtering.sgs}
    \vec{\tau}^{(j)} = \vec{f}^{(j)} \left( \ave{\vec{q}} \right) - \ave{ \vec{f}^{(j)} } \left( \vec{q} \right).
\end{equation}
The product rule in \eqn~\eqref{eq:closure.filtering.product} allows these terms to be explicitly computed.

If the terms in the fluxes $\vec{f}^{(j)}$ were simply related to the conserved variables then the procedure would be complete. However, many of the matter terms, such as the velocities and Lorentz factors, do not have this form. For these other matter variables, generalisations of the product rule~\eqref{eq:closure.filtering.product} are needed on a case-by-case basis.

In common with the closure relations in \sect~\ref{sec:closure.smagorinsky} covariance is explicitly broken as the filtering is applied to a spatial slice. In contrast to the closure relations in \sect~\ref{sec:closure.smagorinsky}, there are no tuneable parameters except for the form of the filtering kernel. Typically the kernel will depend on a lengthscale $L$, but this is usually linked to the numerical grid spacing \gls{dx}, not the physical lengthscales.

\section{GRLES Results}\label{sec:grles}

\subsection{Angular momentum transport in RMNSs}\label{sec:grles.angmom}
\begin{figure}
    \centering
    \includegraphics[width=0.6\textwidth]{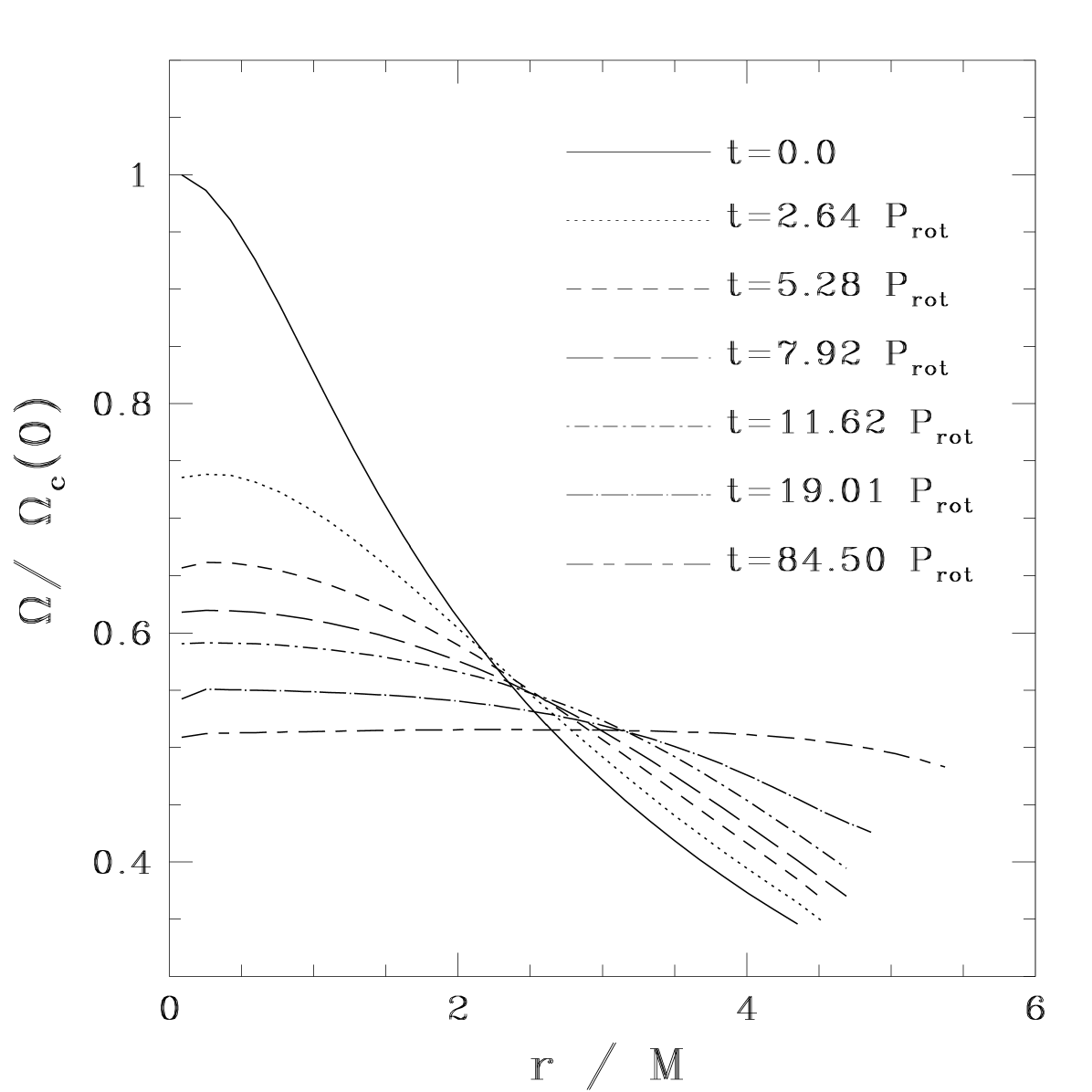}
    \caption{Angular velocity in the equatorial plane for a star evolving under the action of viscosity. The time is given in terms of $P_{{\rm rot}} = 2 \pi / \Omega_c$. The viscous timescale for this model is $\tau_{\rm vis} \simeq 5.5\ P_{\rm rot}$. From Duez et al., Phys.~Rev.~D \textbf{69}:104030 (2004). Reproduced with permission.}
    \label{fig:duez_fig_4}
\end{figure}

\begin{figure}
    \centering
    \includegraphics[width=\textwidth]{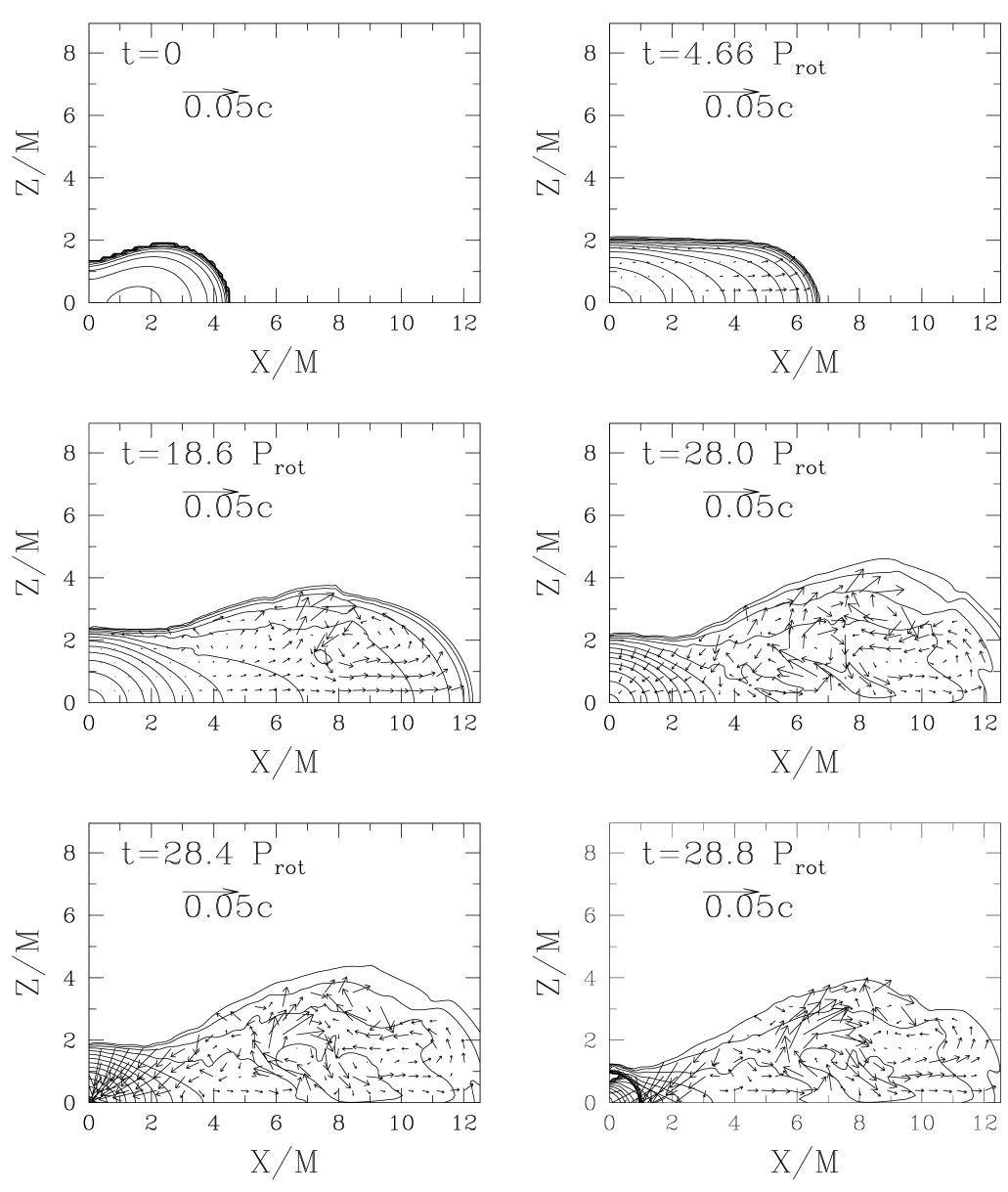}
    \caption{Contours of the rest mass density and velocity profile in the meridional plane for a differentially rotating stars evolving under the action of viscosity. The time is given in terms of $P_{\rm rot} = 2 \pi / \Omega_c$. The dark line in the lower right panel denotes the location of the apparent horizon of the formed \ac{BH}. From Duez et al., Phys.~Rev.~D \textbf{69}:104030 (2004). Reproduced with permission.}
    \label{fig:duez_fig_9}
\end{figure}

The impact of angular momentum transport in \acp{RMNS} was first investigated by \citet{Duez2004}. They studied the evolution of differentially rotating \acp{NS} with a 2D axisymmetric code that solved the relativistic Navier-Stokes equations. In other words, these modelled turbulence using the equations for a viscous flow. This approach is conceptually similar to the Smagorinsky closure of Sec.~\ref{sec:closure.smagorinsky}. However, an important caveat is that, in relativity, the Navier-Stokes equations are mathematically ill-posed \citep{hiscock_generic_1985, kostadt_causality_2000}, so their solution can manifest unphysical instabilities, if these are not numerically suppressed. The initial angular velocity profile was chosen according to the so-called ``\gls{j-specific}-constant'' differential rotation law~\citep{eriguchi_general_1985}. See the $t=0$ line in \fig~\ref{fig:duez_fig_4} for an illustration of the typical angular velocity profiles they considered. They found that these \acp{NS} are driven towards solid body rotation within a few viscous timescales. This is expected, because uniformly rotating configurations are the minimum energy equilibrium configuration for stars of fixed rest mass \gls{Mb-total} and angular momentum \gls{J-total} \citep{hartle_variational_1967}. This process is illustrated in \fig~\ref{fig:duez_fig_4}, which shows the profile of the angular velocity \gls{Omega_ang} in the equatorial plane for selected times. Configurations with total mass above the maximum limit for rigidly rotating stars underwent catastrophic collapse to form \acp{BH} surrounded by massive accretion disks (see \fig~\ref{fig:duez_fig_9}). Configurations with mass below this limit were found to settle to a stationary configuration. 

As already mentioned, the mathematical formalism used in this first work was problematic, because the relativistic Navier-Stokes equations do not admit a well posed initial value problem \citep{hiscock_generic_1985, kostadt_causality_2000}. However, the qualitative findings in \citet{Duez2004} were later confirmed with \ac{ILES} simulations in which viscosity emerged self-consistently from physical \ac{MHD} stresses \citep{duez_evolution_2006, duez_collapse_2006, siegel_magnetorotational_2013}, in simulations employing a variant of the Israel-Stewart hyperbolic formulation of dissipative relativistic \ac{HD} \citep{shibata_general_2017}, and in \ac{GRLES} calculations with a Smagorinsky closure \eqref{eq:smagorinsky} \citep{radice_binary_2018, duez_comparison_2020}. All these studies confirmed that viscosity drives differentially rotating compact objects towards solid body rotation, although there are quantitative differences in the final rotational profiles between the various approaches \citep{duez_comparison_2020}. Several of these studies also confirmed that configurations with mass larger than the rigidly-rotating limit collapse to \acp{BH} over a few viscous timescales, while lower mass objects are stable.

\begin{figure}
    \centering
    \includegraphics[width=0.6\textwidth]{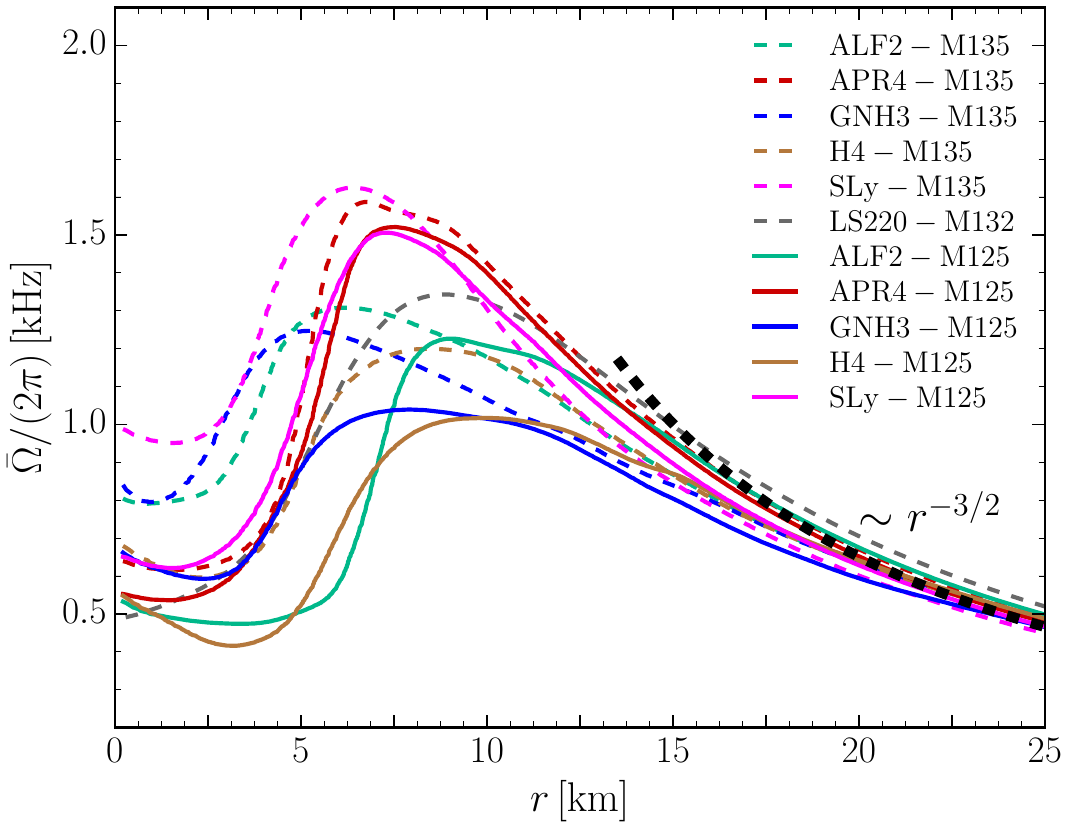}
    \caption{Azimuthally averaged angular velocity profiles for \ac{RMNS} produced by binary mergers with different equations of state. From Hanauske et al., Phys.~Rev.~D \textbf{96}:043004 (2016). Reproduced with permission.}
    \label{fig:hanauske_fig_12}
\end{figure}

That said, while these findings are robust, their applicability to \ac{BNS} mergers is unclear. The reason is that the rotational profiles of \acp{RMNS} are qualitatively different from those predicted by 
the $\gls{j-specific}$-constant law. Rotating stars with \gls{j-specific}-constant profiles have a maximum in the angular velocity at the centre. Centrifugal support is key to support their cores. \acp{RMNS}, instead, have slowly rotating inner cores surrounded by envelopes that are almost entirely supported by rotation (contrast \figs~\ref{fig:hanauske_fig_12} and \ref{fig:duez_fig_4}). This difference was first seen in \citet{shibata_merger_2005} and then studied in more detail in \citet{kastaun_properties_2015, hanauske_rotational_2017, kastaun_structure_2017, ciolfi_general_2017}.

\begin{figure}
    \centering
    \includegraphics[width=0.49\textwidth]{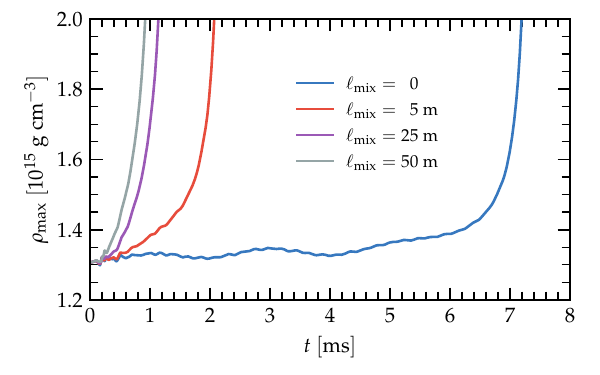}
    \hfill
    \includegraphics[width=0.49\textwidth]{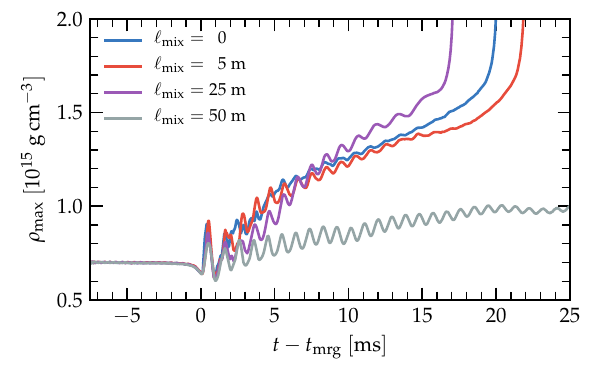}
    \caption{Evolution of the maximum rest mass density for GRLES simulations of $j$-constant differentially rotating stars (left panel) and \ac{BNS} remnants (right panel). Different colors corresponds to different mixing length values \gls{lmix} ranging from $0$ (no subgrid viscosity) to $\SI{50}{\metre}$ (largest subgrid viscosity). Adapted from Radice, Astrophys.~J.~Lett.~\textbf{838}:L2 (2017). Reproduced with permission.}
    \label{fig:radice2017}
\end{figure}

Because of this difference, the outcome of the viscous evolution of \ac{RMNS} is not obvious. This is evident from a comparison of the left and right panels of \fig~\ref{fig:radice2017}. While viscosity is found to monotonically decrease the life time of hypermassive differentially rotating \acp{NS} with \gls{j-specific}-constant rotation law, the impact on a \ac{RMNS} is more complex and large viscosity is found to \emph{delay} the onset of collapse \citep{radice_general-relativistic_2017}. The counter intuitive behavior for large viscosities is due to two effects. First, since the angular velocity has actually a minimum at the center, viscosity can transport angular momentum \emph{into} the remnant. Second, if the postmerger viscous timescale is shorter than the typical \ac{GW} emission timescale ($\gls{J-total}/\gls{J-total-dot} \sim \SI{20}{\milli\second}$; \citealt{bernuzzi_how_2016}), then viscosity can suppress the \ac{GW} emission \citep{shibata_gravitational_2017, radice_general-relativistic_2017}. The result is to increase the total amount of angular momentum of the remnant, which would have otherwise been radiated in \acp{GW}. On the other hand, as we discuss in Sec.~\ref{sec:grles.mhd}, current \ac{GRMHD} models suggest that the viscous time might not be sufficiently short to impact the \ac{GW} emission in a qualitative way. Another reason to exclude the most extreme model shown in \fig~\ref{fig:radice2017} is that large viscosity would produce a massive mildly-relativistic outflow from the thermalization of the tidal streams of the stars shortly before merger \citep{radice_viscous-dynamical_2018}. This ejecta is expected to produce bright radio emission when interacting with the interstellar medium on a timescale of months to years. However, such a scenario is in tension with the non-detection of late-time radio emission from \acp{SGRB} \citep{schroeder_late-time_2020}.

In the more likely scenario in which viscous effects become important only after the remnant has become roughly axisymmetric, viscosity could still either favour or disfavour the collapse. On the one hand, viscosity could drive accretion of the envelope onto the core, thereby triggering its collapse. On the other hand, viscosity could prompt mass ejection and/or transport angular momentum \emph{into} the core of the \ac{RMNS}, thereby increasing its centrifugal support. Even in the case of systems with total mass below the uniformly rotating limit, the effect of viscosity is not fully understood. This is because \acp{RMNS} have angular momentum in significant excess of the maximum that can be supported by uniformly rotating configurations \citep{radice_long-lived_2018}. As such, there is no rigidly-rotating equilibrium configuration towards which systems can evolve while conserving mass and angular momentum. In these cases, the formation of rigidly rotating equilibria is likely preceded by a phase of mass ejection, necessary to remove angular momentum from the system \citep{radice_long-lived_2018}. All these considerations also ignore thermal effects, which can also have either a positive or negative impact on the stability of the remnant \citep{radice_long-lived_2018,hammond_thermal_2021}. 

\begin{figure}
    \centering
    \includegraphics[width=0.6\textwidth]{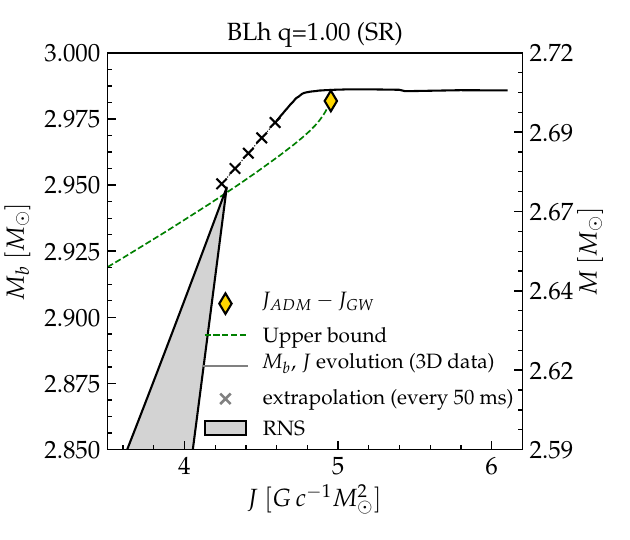}
    \caption{Trajectory of a \ac{RMNS} simulated with the \ac{GRLES} method in the angular momentum ($\gls{J-total}$) and rest-mass ($\gls{Mb-total}$) plane. The diamond denotes the location of the system at the time at which the \ac{GW} timescale $\gls{J-total}/\gls{J-total-dot}$ becomes much larger than the viscous and cooling timescales. The green line shows the trajectory predicted with a simple analytic model, while the crosses are extrapolations in time past the end of the simulation. The grey shaded area is the set of all uniformly rotating equilibria. Depending on the outcome of the viscous evolution, this system might collapse to a \ac{BH}, or end up on one of the stable equilibria in the grey region. From Nedora et al., Astrophys.~J.~\textbf{906}:98 (2021). Reproduced with permission.}
    \label{fig:nedora_fig_3}
\end{figure}

For concreteness, we show in \fig~\ref{fig:nedora_fig_3} an example of a binary highlighting some of the issues discussed so far. The figure shows the evolutionary track of the total angular momentum \gls{J-total} and rest mass \gls{Mb-total} of this particular binary. The binary has a total mass above the maximum limit for a uniformly rotating star, however angular momentum transport operated by viscosity (modelled using the \ac{GRLES} formalism) and hydrodynamic torques of the remnant on the disk drive significant mass outflow from this binary \citep{nedora_spiral-wave_2019,nedora_numerical_2021}. A naive extrapolation of the simulation data in time, as well as a simplistic analytic estimate based on angular momentum conservation, suggests that the remnant could settle into a stable configuration. However, the ultimate outcome of the evolution is unknown. The binary could not be evolved until collapse, or until the end of its viscous phase. This was in part because of the large computational costs, but also, and more importantly, because the neutrino scheme used in this simulation was not adequate to capture the diffusion of radiation over a timescale comparable to the neutrino cooling timescale ($\mathcal{O}(\SI{1}{\second})$). Longer simulations, with sophisticated neutrino transport and viscosity, will be required to clarify these issues.

\subsection{Magnetic field amplification}\label{sec:grles.mhd}
\begin{figure}
\begin{center}
    \includegraphics[width=\textwidth]{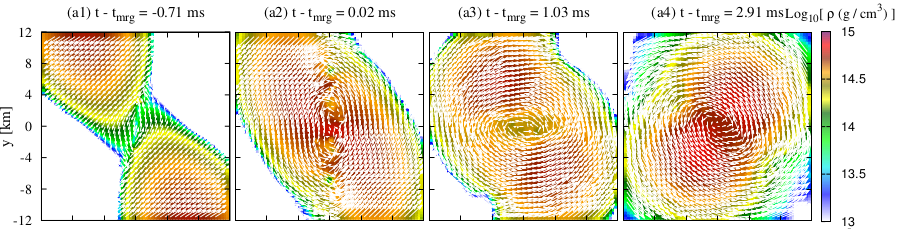}
\end{center}
\caption{\label{fig:kiuchi_2015_fig1} Profile of density and velocity field on the orbital plane for a magnetized BNS merger simulation with a maximum resolution of $\gls{dx} = \SI{17.5}{\metre}$. Adapted from Kiuchi et al., Phys.~Rev.~D \textbf{92}:124034 (2015). Reproduced with permission.} 
\end{figure}

In a seminal paper, \citet{price_producing_2006} first suggested that in \ac{BNS} mergers even weak initial magnetic fields could be amplified by the \ac{KH} instability to magnetar level strengths (${\sim} \SI{e15}{\Gauss}$). They employed Newtonian \ac{SPH} simulations with no explicit turbulence model and their results could not immediately be confirmed by \ac{GR} simulations. This stimulated a vigorous effort in the community to better understand the impact of this instability in mergers. The fact that the \ac{KH} instability is present was investigated in numerical relativity calculations with ILES \citep{anderson_magnetized_2008, baiotti_accurate_2008}, but the large field amplification could not be immediately replicated \citep{giacomazzo_accurate_2011}. We now know that this was because of the insufficient grid resolution of the simulations. Local \ac{ILES} simulations by \citet{obergaulinger_local_2010} and \citet{zrake_magnetic_2013} found that a substantial fraction of the kinetic energy in the shear layer could be converted into magnetic energy during the merger. We remark that the conversion of $10\%$ of the kinetic energy in the shear layer to magnetic energy would be sufficient to produce fields of up to $\SI{e17}{\Gauss}$. The production of magnetar-level fields is thus plausible. The generation of ultra-strong magnetic fields in mergers was finally confirmed by \citet{kiuchi_high_2015, kiuchi_efficient_2015, kiuchi_global_2018}, who performed extremely high-resolution, global \ac{GRMHD} \ac{ILES} simulations of \ac{BNS} mergers. The development of the \ac{KH} instability in one of the simulations of \citet{kiuchi_efficient_2015} is reproduced in \fig~\ref{fig:kiuchi_2015_fig1}.

\begin{figure}
    \centering
    \includegraphics[width=0.6\textwidth]{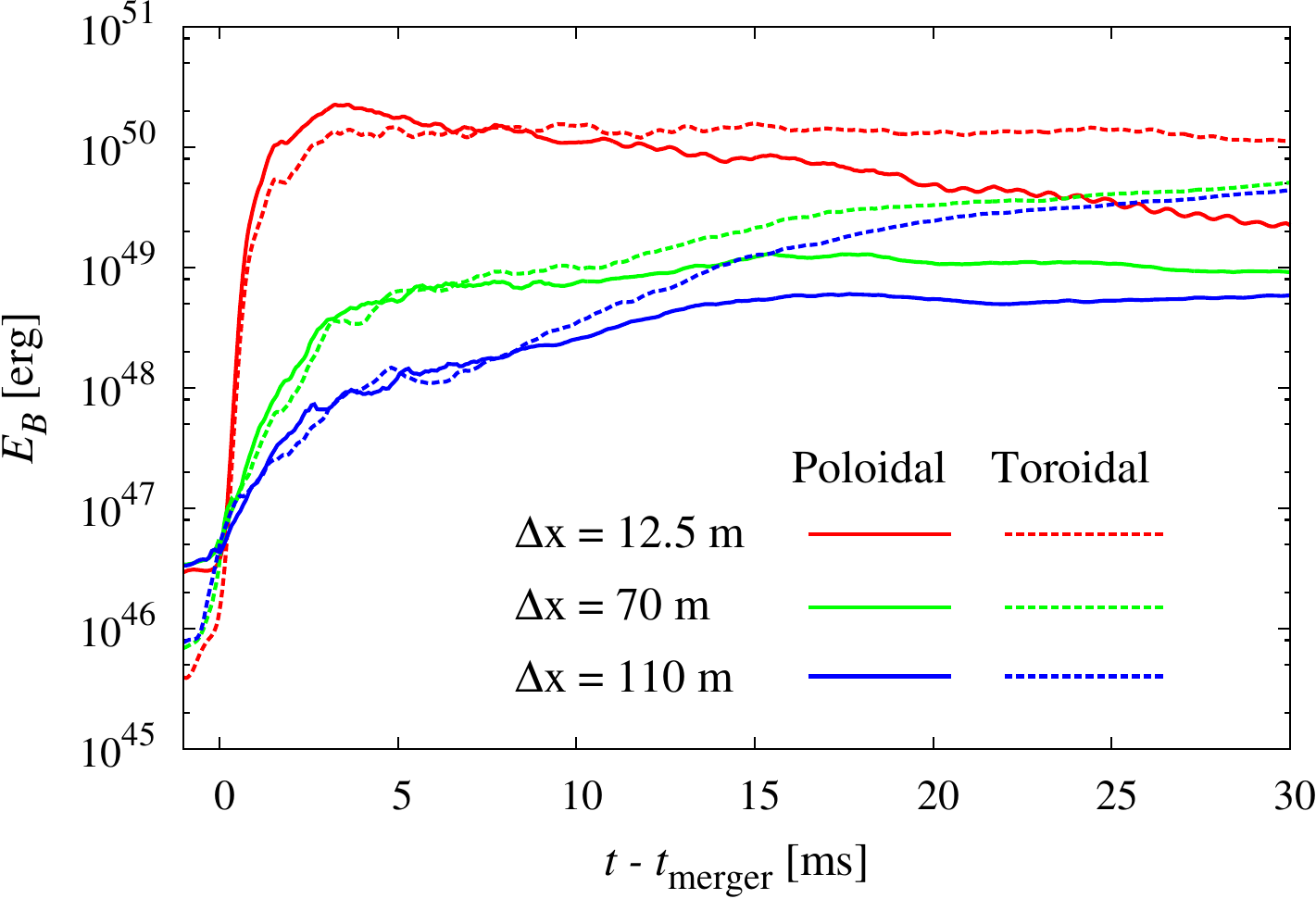}
    \caption{Growth of the magnetic energy due to the KH instability in magnetized BNS simulations with increasing resolution. Adapted from Kiuchi et al., Phys.~Rev.~D \textbf{97}:124039 (2018). Reproduced with permission.}
    \label{fig:kiuchi_2018_fig4}
\end{figure}

Despite the unprecedented, and so far unmatched, grid resolutions, the simulations of Kiuchi et al.~did not achieve convergence. Instead, the saturation strength of the magnetic field after the \ac{KHI} was found to increase monotonically with resolution. The ``low'' resolution calculations, which still had resolution higher than most published simulations, only showed modest amplification of the magnetic field, consistent with those of previous studies \citep[e.g.,][]{giacomazzo_accurate_2011}. As the resolution was increased, both the maximum magnetic field and the growth rate were found to increase, with no sign of saturation (see \fig~\ref{fig:kiuchi_2018_fig4}). They reported fields with strength of up to $\SI{e16}{\Gauss}$. The lack of convergence is not surprising. As we have discussed in Sec.~\ref{sec:general.rmns}, even with a field as large as $\SI{e16}{\Gauss}$, the back reaction of the magnetic field on the fluid is expected to interrupt the normal hydrodynamic cascade only on scales of centimeters, well beyond the resolvable scales in the simulations. For an infinitesimally thin shear layer, the fastest growing mode of the \ac{KHI} is infinitesimal \citep{biskamp_magnetohydrodynamic_2003} and simulations need to capture the magnetic field backreaction scale in order to be converged. In more realistic cases, the fastest growing mode corresponds to the width of the shear layer \citep{biskamp_magnetohydrodynamic_2003}. In the context of binary \ac{NS} mergers the relevant length scale is likely to be the density scale height in the outer layers the \acp{NS} $\gls{rho}/\partial_r \gls{rho} \sim \SI{100}{\meter}$ so it might be possible to fully resolve the linear phase of the \ac{KHI} at extremely high resolutions ($\gls{dx} \sim \SI{5}{\meter}$).

\begin{figure}
    \centering
    \includegraphics[width=0.6\textwidth]{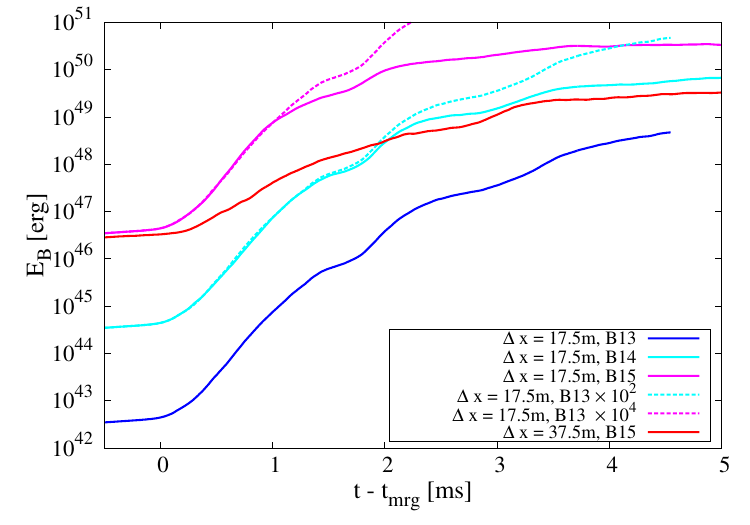}
    \caption{Magnetic field amplification due to the KH instability in high-resolution magnetized BNS simulations with different initial magnetic field strengths. Adapted from Kiuchi et al., Phys.~Rev.~D \textbf{92}:124034 (2015). Reproduced with permission.}
    \label{fig:kiuchi_2015_fig5}
\end{figure}

One might wonder whether the stringent resolution requirements for the \ac{KH} instability can be bypassed by starting simulations with initially large (${\sim} \SI{e16}{\Gauss}$) magnetic fields. Unfortunately, the results of \citet{kiuchi_efficient_2015} do not support this practice. Indeed, as shown in \fig~\ref{fig:kiuchi_2015_fig5}, they found substantial differences between results of simulations employing initially large fields and re-scaled results from calculations that started with more realistic initial field strengths. Such differences do not manifest themselves in the initial growth of the \ac{KH} instability, when the magnetic field back reaction is unresolved, but they are apparent in the subsequent evolution. These results suggest that nonlinear \ac{MHD} effects are important after the merger and that realistic initial field values are required for the simulations to be predictive. This point is also stressed by \citet{aguilera-miret_role_2023} using results from \ac{GRLES} simulations with the gradient expansion method discussed in \ref{sec:closure.explicit}.

\citet{kiuchi_global_2018} studied the role of the \ac{MRI} and of magnetic stresses on the evolution of the \acp{RMNS} over a timescale of $\SI{30}{\milli\second}$ after the merger using \ac{ILES} simulations. In particular, they measured the effective alpha viscosity induced by magnetic stresses. They found that magnetic stresses are most important in the rotationally supported envelope of the \ac{RMNS}, peaking at densities $\gls{rho} < \SI{e13}{\gram\per\centi\metre\cubed}$, where $\gls{alpha-viscosity} \sim 0.01{-}0.02$. The viscosity at higher densities was found to be significantly smaller, $\gls{alpha-viscosity} \sim 0.0005{-}0.005$. Although they cautioned that their results might not be yet converged, the reported values of $\gls{alpha-viscosity}$ at $\gls{rho} \sim \SI{e13}{\gram\per\centi\metre\cubed}$ for different resolutions appear to be converging at first order to ${\sim}0.005$. On the basis of these results, the impact of magnetic torques on the stability of the \ac{RMNS} and on its \ac{GW} signature could be expected to only manifest over very long timescales. 

These results are consistent with the fact that the region with $\gls{rho} > \SI{e13}{\gram\per\centi\metre\cubed}$ is stable against the \ac{MRI}, since $\dv{\gls{Omega_ang}}{r} > 0$
(see \fig~\ref{fig:hanauske_fig_12}). \citet{radice_binary_2020} and \citet{radice_ab-initio_2023} used the values of the viscosity measured by \citet{kiuchi_global_2018} and \citet{kiuchi_large-scale_2023} to calibrate their Smagorinsky-type subgrid model. They confirmed that the turbulent viscosity does not have a major impact on the stability of the \ac{RMNS}. In particular, while differences in the \ac{GW} amplitude were recorded, the post-merger \ac{GW} peak-frequency was found to be unaffected.

Convergent results of the magnetic field amplification due to the \ac{KH} instability could only be achieved with the introduction of subgrid models. \citet{giacomazzo_producing_2015} pioneered this approach with a subgrid model inspired by mean-field dynamo theory. Their model reproduced the quick amplification of weak fields to more than $\SI{e16}{\Gauss}$. Their simulation showed consistent results across resolutions, in contrast with the direct simulations of \citet{kiuchi_high_2015, kiuchi_efficient_2015, kiuchi_global_2018}. However, their subgrid model also predicted large field amplification at the surface of the \acp{NS}, driven by the large, unphysical velocity gradients present in the artificial atmosphere of their simulations. As such, while promising, their approach still required fine tuning of the model parameters and it was not fully predictive. A similar mean-field dynamo model has also been recently employed by \citet{most_flares_2023, most_impact_2023} to study the production of flares from long-lived \ac{RMNS}.

\begin{figure}
    \centering
    \includegraphics[width=0.6\textwidth]{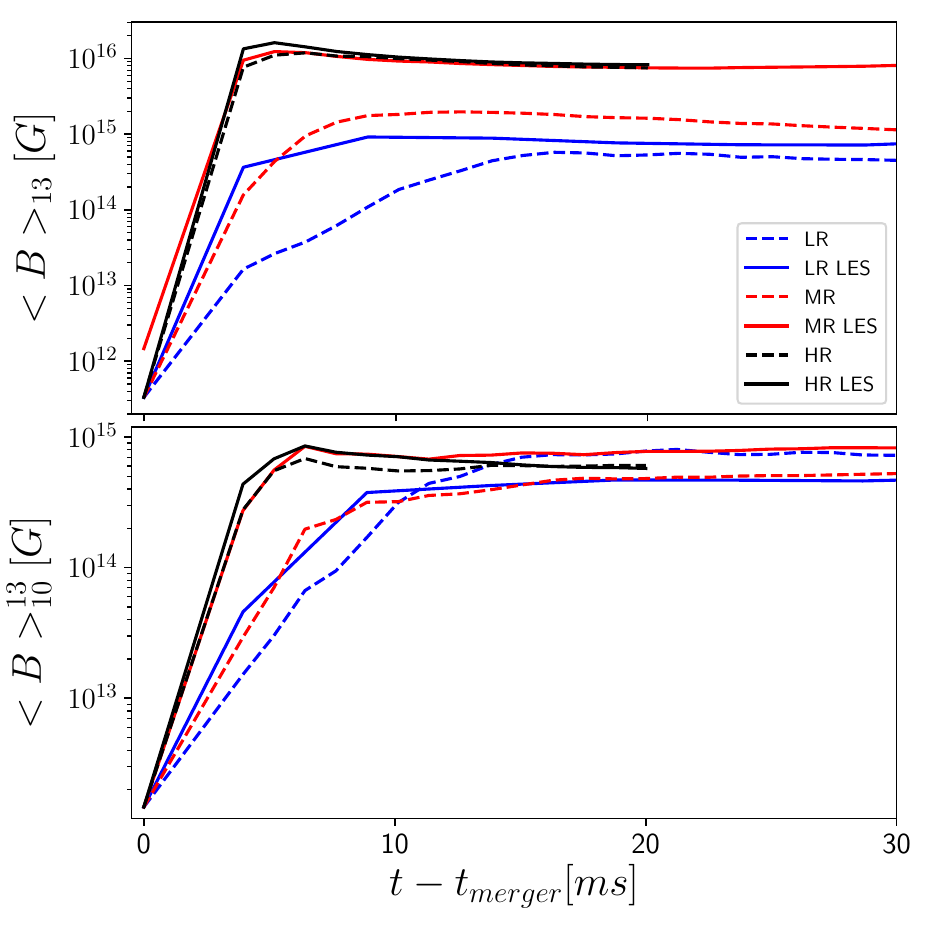}
    \caption{Evolution of the average magnetic field strength in GRLES simulations of magnetized BNS mergers. The figure shows the magnetic field in the bulk of the remnant ($\gls{rho} > \SI{e13}{\gram\per\centi\metre\cubed}$; top panel) and in the envelope ($\SI{e10}{\gram\per\centi\metre\cubed} < \gls{rho} < \SI{e13}{\gram\per\centi\metre\cubed}$; bottom panel). The dashed lines are for simulations with no subgrid models, while the solid lines show the results obtained with a subgrid model. The different resolutions are $\gls{dx} = \SI{120}{\metre}$ (LR), $\gls{dx} = \SI{60}{\metre}$ (MR), and $\gls{dx} = \SI{30}{\metre}$ (HR). From Palenzuela et al., Phys.~Rev.~D \textbf{106}:023013 (2022). Reproduced with permission.}
    \label{fig:palenzuela_2022_fig5}
\end{figure}

\citet{aguilera-miret_universality_2022} and \citet{palenzuela_turbulent_2022} presented the first simulations with a gradient subgrid model that had no tunable parameters, except for the adopted filter function as outlined in \sect~\ref{sec:closure.explicit}. Their simulations confirmed that the \ac{KH} instability can amplify weak magnetic fields in the stars prior to merger to magnetar-level fields $\SI{e16}{\Gauss}$. The statistical properties of the fields appear to have converged \citep{palenzuela_turbulent_2022}. For example, \fig~\ref{fig:palenzuela_2022_fig5}, adapted from \citet{palenzuela_turbulent_2022}, shows the average magnetic field strength in different region of the \ac{RMNS}. As in \citet{kiuchi_global_2018}, the maximum strength and growth rate of the magnetic field were found to initially increase as the grid spacing was reduced. However, and in contrast to the \ac{ILES} simulations of \citet{kiuchi_global_2018}, the GRLES simulations appear to achieve convergence at the highest resolutions.

\begin{figure}
    \centering
    \includegraphics[width=\textwidth]{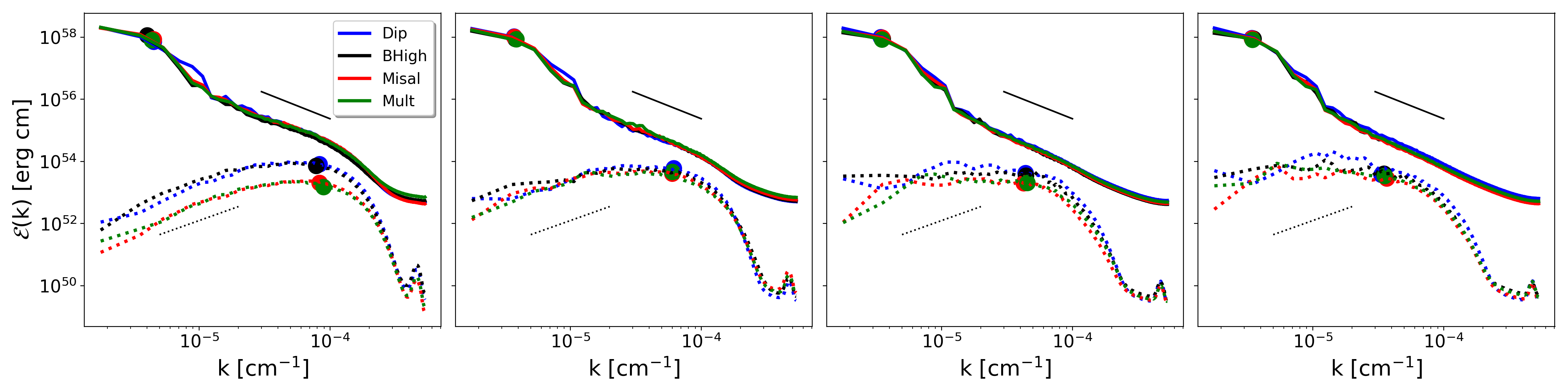}
    \caption{Kinetic (solid) and magnetic (dotted) energy spectra from GRLES simulations of magnetized BNS mergers. From left to right, $t - t_{\rm mrg} = \SI{5}{\milli\second}, \SI{10}{\milli\second}, \SI{20}{\milli\second}, \SI{30}{\milli\second}$. The different colors correspond to different initial magnetic field configurations, see \citet{aguilera-miret_universality_2022} for the details. The dots show the integral scales for the kinetic and magnetic spectra. From Aguilera-Miret et al., Astrophys.~J.~Lett.~\textbf{926}:L31 (2022) 2. Reproduced with permission.}
    \label{fig:aguilera_miret_2021_fig4}
\end{figure}

The maximum strength and ratio of toroidal to poloidal field components shortly after merger was found to be independent of the initial magnetic field in the stars, as long as that was sufficiently weak for the back reaction on the fluid, at the characteristic scale of the eddies generated by the \ac{KH} instability (${\sim} \SI{1}{\kilo\metre}$; Sec.~\ref{sec:ns_conditions}), to be negligible \citep{aguilera-miret_universality_2022}. This is illustrated by \fig~\ref{fig:aguilera_miret_2021_fig4}, which compares kinetic and magnetic energy spectra obtained with different initial magnetic field configurations. These result are consistent with the expectations for a turbulent dynamo. An important implication is that the poor knowledge of the interior structure of the magnetic field in \acp{NS} prior to merger, \citep[e.g.,][]{braithwaite_evolution_2006, lasky_hydromagnetic_2011, ciolfi_instability-driven_2011, bilous_nicer_2019, sur_long-term_2022}, might not compromise the predictive power of simulations.

\begin{figure}
    \centering
    \includegraphics[width=\textwidth]{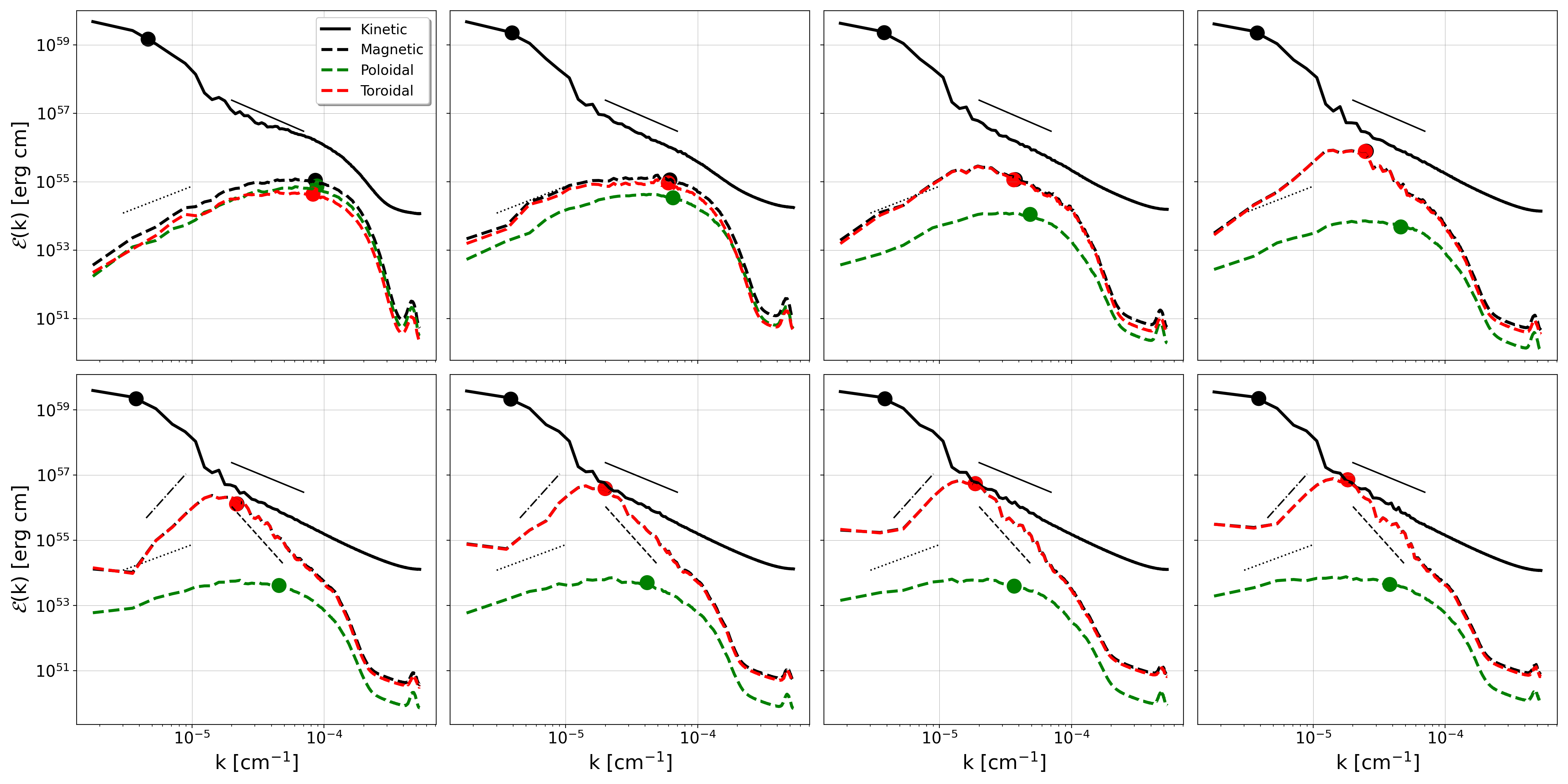}
    \caption{Kinetic (solid) and magnetic (dotted) energy spectra from GRLES simulations of magnetized BNS mergers. From left to right and top to bottom, $t - t_{\rm mrg} = \SI{5}{\milli\second}, \SI{10}{\milli\second}, \SI{20}{\milli\second}, \SI{30}{\milli\second},  \SI{50}{\milli\second},  \SI{80}{\milli\second},  \SI{100}{\milli\second},  \SI{110}{\milli\second}$. The dots show the integral scales for the kinetic and magnetic spectra. From Aguilera-Miret et al., Phys.~Rev.~D \textbf{108}:103001 (2023). Reproduced with permission.}
    \label{fig:aguilera_miret_2023_fig10}
\end{figure}

\begin{figure}
    \centering
    \includegraphics[width=0.6\textwidth]{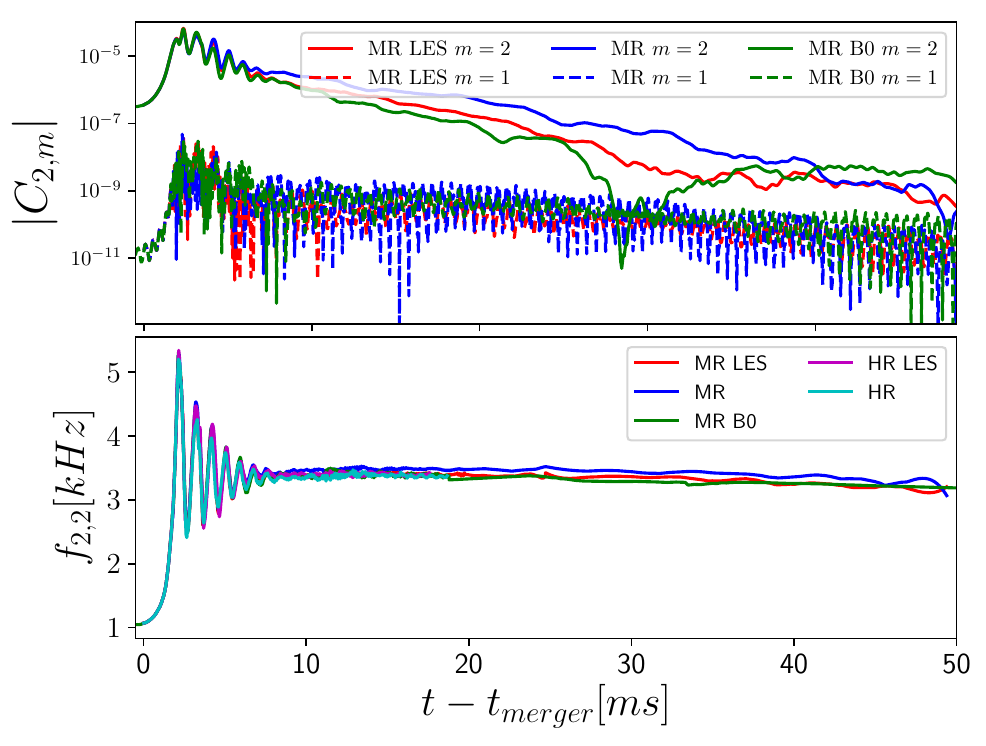}
    \caption{Curvature GW signal ($\Psi_4$) from GRLES simulations of magnetized BNS mergers. Upper panel: amplitude of the $\ell=2, m=2$ and $\ell=2, m = 1$ modes. Lower panel: instantaneous frequency of the $\ell = 2, m = 2$ mode. The dashed lines are for simulations with no subgrid models, while the solid lines show the results obtained with a subgrid model. From Palenzuela et al., Phys.~Rev.~D \textbf{106}:023013 (2022). Reproduced with permission.}
    \label{fig:palenzuela_2022_fig17}
\end{figure}

\citet{aguilera-miret_universality_2022} and \citet{aguilera-miret_role_2023} also found tantalizing evidence for an inverse cascade of the magnetic field. In particular, they reported a growth of the typical magnetic field scale from ${\sim} \SI{500}{\metre}$, immediately after merger, to ${\sim}\SI{3.5}{\kilo\metre}$, $\SI{110}{\milli\second}$ after merger. This is illustrated in \fig~\ref{fig:aguilera_miret_2023_fig10}, reproduced from their second work, which shows the growth of the field scales. It can also be noticed that the magnetic field reaches equipartition with the kinetic energy at its integral length scale, as expected from dynamo theory. However, the field remained predominantly toroidal and no significant angular momentum or viscous effects, which would be mediated by a poloidal field, were reported \citep{aguilera-miret_universality_2022}. \citet{palenzuela_turbulent_2022} also discussed the simulations presented by \citet{aguilera-miret_universality_2022}. They reported only modest differences in the amplitude of the \ac{GW} signal between simulations that resolved the growth of the magnetic fields and those with small post-merger field, see \fig~\ref{fig:palenzuela_2022_fig17}. These studies are currently being extended to include neutrino cooling and heating effects \citep{palenzuela_large_2022, miravet-tenes_assessment_2022}. The changes in the gravitational wave peak frequency were even smaller. This suggests that it will be possible to control for uncertainties due to turbulence in future constraints on the dense matter \ac{EOS} using the postmerger \ac{GW} spectrum \citep{bauswein_exploring_2016, bernuzzi_neutron_2020}.

More recently, \citet{kiuchi_large-scale_2023} presented results of ILES GRMHD simulations with approximate neutrino-transport at extremely high-resolution (with $\gls{dx}$ as small as $\SI{12.5}{\metre}$) and found compelling evidence for an $\alpha\Omega$-dynamo \citep{Brandenburg2005} operating in the shear layer at the interface between the star and the disk. Accordingly, poloidal flux is generated from the toroidal field by the turbulent flow (the $\alpha$-effect) and then converted into toroidal field by the rotation (the $\Omega$-effect). However, \citet{kiuchi_large-scale_2023} had to employ an artificially large initial field (\SI{e15.5}{\Gauss}), in order to resolve the \ac{MRI} in this region. As such, even though these results suggests that there is an inverse turbulent cascade, similar to those reported by \citet{aguilera-miret_universality_2022, aguilera-miret_role_2023}, it is not clear that these studies are describing the same physics. Indeed, the simulations of \citet{aguilera-miret_universality_2022, aguilera-miret_role_2023} show an inverse cascade for the toroidal field, while those of \citet{kiuchi_large-scale_2023} shows a dynamo producing a poloidal field. Overall, the results of Kiuchi and collaborators suggest that an $\alpha\Omega$-dynamo is likely to operate in the remnant, but its implications for the dynamics and multi-messenger signals still need to be understood.

\section{Future directions}\label{sec:future}

Turbulence plays an important role in the dynamics of \ac{NS} mergers. However, due to the enormous separation between the global scale of the system and the dissipation range, direct numerical simulations capturing the full physics of \ac{NS} mergers are unfeasible. The \ac{GRLES} methodology replaces the fundamental equations of \ac{GRMHD} with an effective theory in which the equations are coarse grained. Small scale effects are included in the form of subgrid-scale closures. 

\ac{GRLES} simulations of \ac{NS} mergers have now been performed by several groups. These simulations confirmed that the \ac{KH} instability in binary \ac{NS} mergers generate strong turbulent magnetic fields. A dynamo process involving the turbulent generation of a poloidal field and its conversion into a toroidal field by the differential rotation has been shown to be active in the exterior layers of the \ac{RMNS} \citep{kiuchi_large-scale_2023}. Within this context, turbulent resistivity is needed to rearrange the field lines and create large scale magnetic structures \citep{aguilera-miret_role_2023}. The measured effective viscosity due to magnetic stresses is small. This suggests that turbulence will not jeopardize proposed studies of the dense matter EOS with next-generation \ac{GW} experiments \citep{kiuchi_global_2018, radice_binary_2020, palenzuela_turbulent_2022, breschi_constraints_2022}. However, more studies are needed to fully quantify this statement. On the other hand, it is expected that turbulence and dynamo action will affect the long-term dynamical evolution of the remnant and leave a significant imprint on the associated multi-messenger emission and nucleosynthesis yields.

The \ac{GRLES} methodology is very well developed for non-relativistic flow, particularly in the unmagnetized limit. Its application to the relativistic, magnetized flows developing in \ac{NS} mergers is in its infancy. There are conceptual and practical issues that need to be resolved concerning the \ac{GRLES} method. In particular, we have discussed the issue of covariance of different formulations and their interpretation. Another issue is model validation. Unlike terrestrial cases, experimental testing and validation is completely impractical. Current validation techniques rely on comparison to fine scale simulations. The potential systematic errors arising from whether these fine scale simulations are truly in the \ac{DNS} regime, or whether the models transfer to a different spacetime (for example), are currently unknown. Moreover, it is not obvious which observables a subgrid \ac{GRLES} model should be tuned to. It is now standard for models to be validated against the statistics of a (suitably filtered) \ac{DNS} model. However, it is typical to validate using statistics of the strain-rate or energy transfer. In the astrophysical context these are unobservable; instead we are interested in observables such as the \ac{GW} signal or the neutrino lightcurves. Quantifying the systematics from the difference between the validated and observed statistics is necessary future work.

More generally, an underlying problem for the future of \ac{NS} merger simulations is uncertainty quantification. The many uncertainties within the models include the \ac{EOS} and transport coefficients and aspects of the initial data. There are formal techniques for studying how this uncertainty propagates to impact on observable quantities (reviewed, for example, by~\citealt{abgrall_uncertainty_2017}). However, these techniques suffer from the curse of dimensionality: the high dimension of the parameter space makes the quantification uncertain and inaccurate. The turbulent closures introduced by \ac{GRLES} methods are necessary but introduce many more parameters. This makes a rigorous understanding of the accuracy with which parameters can be estimated from, for example, observed \ac{GW} signals, much more complex. 

We anticipate that in the next few years there will be substantial progress in our understanding of turbulence in \ac{NS} mergers. Work is currently ongoing on three fronts 1) developing simulations that combine sophisticated microphysics, \ac{MHD}, and \ac{GRLES}, \citep[e.g.,][]{palenzuela_large_2022, zappa_binary_2023}; 2) improvement in phenomenological subgrid models using local simulations, \citep[e.g.,][]{miravet-tenes_assessment_2022, miravet-tenes_assessment_2023}; and 3) development of model agnostic, data-driven, subgrid models, \citep[e.g.,][]{brunton_machine_2020, yuan_deconvolutional_2020, karpov_physics-informed_2022}.

\backmatter

\end{document}